%Paper: astro-ph/9311053
%From: max@physics.Berkeley.EDU (Max Tegmark)
%Date: Fri, 19 Nov 93 10:07:33 PST
%Date (revised): Sun, 19 Dec 93 10:26:55 PST
%Date (revised): Sun, 19 Dec 93 10:36:44 PST

% This file is in plain TeX.
% Postscript files containing the figures are in the
% accompanying email, tar-compressed and
% uuencoded. Do the following to print the paper with the
% figures:
% * Save the email message accompanying this one, say as "figures.uu".
% * Remove the mail-header junk.
% * Type
%   csh figures.uu
%   and the file will unpack itself automatically.
% * Save the the file you are currently reading, say as
    'constraints.tex'.
% * Delete the line reading
% \end    %HELLO! TO PRINT THE FIGURES AS WELL, DELETE THIS LINE.
%   from the middle of this TeX file.
% * tex constraints.tex
% * dvips constraints.tex
%
%       :-)
% =========================================================

% ---------------------------------
% JUNK FOR NICE COVER PAGE

\font\titlefont=cmb10 at 15truept
\font\namefont=cmr12 at 14truept
\font\addrfont=cmti12
\font\rmtwelve=cmr12

% ABSTRACTS:
\newbox\abstr
\def\abstract#1{\setbox\abstr=
    \vbox{\hsize 5.0truein{\par\noindent#1}}
    \centerline{ABSTRACT} \vskip12pt
    \hbox to \hsize{\hfill\box\abstr\hfill}}

%  macro for invoking today's date (when TeX is run on your file)
\def\today{\ifcase\month\or
        January\or February\or March\or April\or May\or June\or
        July\or August\or September\or October\or November\or December\fi
        \space\number\day, \number\year}

\def\author#1{{\namefont\centerline{#1}}}
\def\addr#1{{\addrfont\centerline{#1}}}

{ % SETTINGS FOR COVER PAGE ONLY
\rmtwelve

%\hoffset=1 truein
%\voffset=1 truein
\vsize=9 truein
\hsize=6.5 truein
%\parskip=0.2truecm
\raggedbottom
\baselineskip=16pt

%\today\hfill CfPA 93-th-40
December 16, 1993\hfill CfPA 93-th-40

\vskip2.5truecm
\centerline
{\titlefont Power Spectrum Independent Constraints}
\centerline
{\titlefont on Cosmological Models
\footnote{$^\dagger$}{\rm Submitted to Ap. J. in November 1993,
revised in December 1993.}}
\nobreak
  \vskip 0.5truecm

  \author{Max Tegmark}
  \smallskip

 \author{Emory Bunn}
  \smallskip

 \author{Wayne Hu}
  \smallskip

  \addr{Department of Physics, University
of California,}
  \addr{Berkeley, California  94720}
  \bigskip
  \bigskip

\abstract{
A formalism is presented that allows cosmological
experiments to be tested for consistency, and allows a simple frequentist
interpretation of the resulting significance levels.
As an example of an application, this formalism is used to place
constraints on bulk flows of galaxies using the
results of the microwave background anisotropy experiments COBE and SP91,
and a few simplifying approximations about the experimental
window functions.
It is found that if taken at face value, with the quoted errors,
the recent detection by Lauer and Postman of a bulk
flow of 689 km/s on scales of 150$h^{-1}$Mpc
is inconsistent with SP91 at a 95\% confidence level
within the framework of a Cold Dark Matter (CDM)
model.
The same consistency test is also used to place constraints that are
completely
model-independent, in the sense that they hold for any
power spectrum whatsoever
--- the only assumption being that the random fields are Gaussian.
It is shown that the resulting infinite-dimensional optimization
problem reduces
to a set of coupled non-linear equations that can readily be solved
numerically.
Applying this technique to the above-mentioned example, we find that the
Lauer and Postman result is inconsistent with SP91
even if no
assumptions whatsoever are made about the power spectrum. }
}  % End of cover settings
\vfill\eject

%
% HERE COMES THE MAIN PART OF THE PAPER:
%

\baselineskip12pt
\hoffset1.46cm
\hsize 15.59cm

%
% EQUATION NUMBERING STUFF:
%
\def\inc#1{\hbox{\global \advance#1 1}}
\countdef\eqnr=201    \countdef\refnr=202
\eqnr=1             \refnr=1
\def\nexteq{\inc{\eqnr}}
\def\enr{\number\eqnr\nexteq}
\def\eq{\eqno(\enr)}

%
% REFERENCE STUFF:
%
\def\pp{\noindent\parshape 2 0truecm 15.6truecm 2truecm 13.6truecm}
\def\rf#1;#2;#3;#4 {\par\pp#1, #2, #3, #4. \par}
\def\rn{\pp}

\countdef\nonlinEq=221
\countdef\ContMinEq=222
\countdef\DiscMinEq=223
\countdef\TedEqOne=231
\countdef\TedEqTwo=232
\countdef\TedEqThree=233
\countdef\TedEqFour=234

\def\llist{\noindent\parshape 2 0.5cm 15.0cm 1.32cm 14.18cm}

\def\blist{\noindent\parshape 2 0.8cm 14.7cm 1.00cm 14.50cm$\bullet$}
\def\streck{\noalign{\vskip2pt\hrule\vskip2pt}}
\def\streck{\noalign{\vskip2pt\hrule\vskip2pt}}
\def\subsection#1{\bigskip\goodbreak{\bf#1}\bigskip}

\def\ie{{\frenchspacing\it i.e.}}

\def\etc{{\it etc.}}
\def\etal{{\it et al.}}

\def\tt#1{\times10^{#1}}

% UNITS:

\def\K{{\rm K}}
\def\s{{\rm s}}

\def\Mpc{{\rm Mpc}}

\def\expec#1{\langle#1\rangle}

\def\crr{\cr\noalign{\vskip2pt}}
\def\Wc{W_{cobe}}
\def\Wsp{W_{sp}}

\def\Wlp{W_{lp}}
\def\clp{c_{lp}} \def\csp{c_{sp}}  \def\ccobe{c_{c}}
\def\slp{s_{lp}} \def\ssp{s_{sp}}  \def\scobe{s_{c}}
\def\slph{\hat{s}_{lp}} \def\ssph{\hat{s}_{sp}}  \def\scobeh{\hat{s}_{c}}
\def\f{f}
 \def\fsp{\f_{sp}}  \def\fcobe{\f_{c}}
\def\vlp{V_{lp}} \def\vsp{V_{sp}}

\def\jbar{\bar j}

\def\crr{\cr\noalign{\vskip 4pt}}
\def\t{\tau_0}

\def\sumi{\sum_{i=1}^n}
\def\sumj{\sum_{j=1}^n}
\def\intk{\int_0^{\infty}}
\def\vl{{\bf\lambda}}
\def\vp{{\bf p}}     
\def\va{{\bf a}} \def\vb{{\bf b}} \def\vc{{\bf c}}
\def\vv{{\bf v}} \def\ve{{\bf \epsilon}}
\def\x{\eta}
\def\xh{\hat\x}
\def\s{s}
\def\sh{\hat s}
\def\c{c}
\def\fx{f_{\x}}  \def\Fx{F_{\x}}

\def\kms{{\,\rm km/s}} \def\mK{{\,\rm \mu K}}
\def\N{N}

\beginsection{1. INTRODUCTION}

Together with the classical cosmological parameters $h$, $\Omega$, {\etc},
the power spectrum $P(k)$ of cosmological density fluctuations is one of
the most sought-for quantities in modern cosmology, vital for
understanding both the formation of large-scale
structure and the fluctuations in the cosmic microwave background radiation
(CMB).

The traditional approach has been to assume some functional
form for $P(k)$
(like that predicted by the cold dark matter (CDM) scenario, for instance),
and then investigate whether the predictions of the model are
consistent with experimental data or not.
The large amounts of data currently being produced by
new CMB experiments
and galaxy surveys, all probing different parts of the power spectrum,
allow a new and more attractive approach. We can now begin to probe exact
shape of the function $P(k)$, without making any prior assumptions about
$P(k)$.
More specifically, we measure different weighted averages of the function, the
weights being the experimental window functions.

This new approach is quite timely (Juskiewicz 1993), as there are now many
indications that the primordial power spectrum may have been more
complicated than an $n=1$ power law.
There are several sources of concern about the standard CDM
cosmology, with inflation leading to $\Omega\approx 1$ and a primordial
$n\approx 1$ Harrison-Zel'dovich power spectrum.
Compared to COBE-normalized CDM, observational data shows
unexpected
large-scale bulk flows  (Lauer \& Postman 1993),
too weak density correlations on small scales (Maddox {\etal} 1990),
a rather quiet local velocity
field (Schlegel {\it et al.} 1993) and a deficit of hot x-ray clusters
(Oukbir \& Blanchard 1992).
The combined data from the COBE DMR (Smoot {\etal} 1992) and the Tenerife
anisotropy experiment (Watson {\etal} 1992) point to a spectral
index exceeding unity (Watson \& Guti\'errez de la Cruz 1993)
which, if correct, cannot be explained by any of the
standard inflationary models.
The recent possible detections of halo gravitational microlensing events
(Alcock
{\etal} 1993) give increased credibility to the possibility that the dark
matter
in our galactic halo may be baryonic. If this is indeed the case, models
with $\Omega<1$ and nothing but
baryonic dark matter (BDM) (Peebles 1987,
Gnedin \& Ostriker 1992,
Cen, Ostriker \& Peebles 1993) become rather appealing.
However, in contrast to CDM with inflation,
BDM models do not include a physical
mechanism that makes a unique prediction for what the primeval power spectrum
should be. Rather, the commonly assumed $P(k)\propto k^{-1/2}$ is
chosen {\it ad hoc} to fit observational data.
Moreover, for fluctuations near the curvature scale in open
universes, where the
$\Omega=1$ Fourier modes are replaced by hyperspherical Bessel functions
with the curvature radius as a built-in length scale,
the whole notion of scale-invariance loses its meaning
(Kamionowski \& Spergel 1993).

In summary, it may be advisable to avoid
theoretical prejudice as to the shape of the primordial
power spectrum.
In this spirit, we will develop a consistency test that
requires no such assumptions whatsoever about the form of the
power spectrum. This approach was pioneered by
Juszkiewitz, G\'orski and Silk (1987), who developed a formalism for
comparing two experiments in a power-spectrum independent manner.
We generalize this method
to the case of more than two experiments, and then use the formalism to
assess the consistency of three recent observational results:  the
CMB anisotropy measurements made by the COBE Differential Microwave Radiometer
(Smoot {\etal} 1992), the South Pole anisotropy experiment (SP91, Gaier {\etal}
1992), and the measurement of bulk velocity of Abell clusters in a 150 $h^{-1}
\,{\rm Mpc}$
sphere (Lauer and Postman 1993, hereafter LP).

In Section 2, we develop a formalism for testing cosmological models for
consistency. In Section 3, we apply this formalism to the special case of cold
dark matter (CDM) and the LP, SP91 and COBE experiments. In Section 4, we solve
the variational problem that arises in consistency tests of models where we
allow arbitrary power-spectra, and apply these results to the LP,
SP91 and COBE experiments. Section 5 contains a discussion of our results.
Finally, the relevant window functions are derived in
the Appendix.

\beginsection{2. CONSISTENCY TESTS FOR COSMOLOGICAL MODELS}

In cosmology, a field where error bars tend
to be large, conclusions can depend crucially on the
probabilistic interpretation of
confidence limits.
Confusion has sometimes arisen
from the fact that large-scale measurements of microwave background
anisotropies and bulk flows are fraught with two quite distinct sources of
statistical uncertainty, usually termed experimental noise and
cosmic variance.
In this section, we present a detailed prescription for testing any
model for consistency with experiments, and discuss the
appropriate probabilistic
interpretation of this test.
By {\it model} we will mean not merely a model for the  underlying
physics, which predicts the physical quantities that we wish to
measure, but also a model for the various experiments.
Such a model is allowed to contain any number of free parameters.
In subsequent sections, we give examples of both a very narrow class of models
(standard CDM where the only free parameter
is the overall normalization of the power spectrum),
and a wider class of models
(gravitational instability with Gaussian adiabatic fluctuations
in a flat universe with the standard recombination history,
the power spectrum being an arbitrary function).

Suppose that we are interested in $\N$ physical quantities
$\c_1, ...,\c_{\N}$, and have $\N$ experiments $E_1,...,E_{\N}$
devised such that the experiment $E_i$ measures the quantity $\c_i$.
Let $\s_i$ denote the number actually obtained by the
experiment $E_i$. Because of experimental noise, cosmic variance,
{\etc}, we do not expect $\s_i$ to exactly equal $\c_i$. Rather,
$\s_i$ is a random variable that will yield different values each
time the experiment is repeated. By repeating the experiment $M$
times on this planet and averaging the results, the uncertainty
due to experimental noise can be reduced by a factor $\sqrt M$.
However, if the same experiment were carried out in a number of
different horizon volumes throughout the universe (or, if we have
ergodicity, in an ensemble of universes with different realizations of
the underlying random field), the results would also be expected to
differ. This second source of uncertainty is known as cosmic variance.
We will treat both of these uncertainties together by simply
requiring the model to specify the probability distribution for the
random variables $\s_i$.

Let us assume that the random variables $s_i$ are all independent, so
that the joint probability distribution is
simply the product of the individual probability distributions, which
we will denote $f_i(s)$.
This is an excellent approximation for the microwave
background and bulk flow experiments we will consider.
Finally, let $\sh_1, ...,\sh_{\N}$ denote the numbers actually obtained in one
realization of the experiments.

The general procedure for statistical testing will be as follows:

\blist{First, define a parameter $\x$ that is some sort of measure of
how well the observed data $\s_i$ agree with the probability
distributions $f_i$, with higher $\x$ corresponding to a better fit.}

\blist{Then compute the probability distribution $f_{\x}(\x)$ of this
parameter, either analytically or by employing Monte-Carlo
techniques.}

\blist{Compute the observed value of $\x$, which we will denote
$\xh$.}

\blist{Finally, compute the probability $P(\x<\xh)$, {\ie} the
probability of getting as bad agreement as we do or worse.}

We will now discuss these four steps in more detail.

\subsection{2.1. Choosing a goodness-of-fit parameter}

Obviously, the ability of to reject models at a high level of
significance depends crucially on making a good choice of
goodness-of-fit parameter $\eta$.
In the literature, a common choice is the {\it likelihood product}, {\ie}
$$\x_l\equiv\prod_{i=1}^{\N} f_i(\s_i).\eq$$
In this paper, we will instead use the {\it probability product}, {\ie} the
product of the
probabilities $P_i$ that each of the experiments yield results at least as
extreme as
observed.
Thus if the observed $\sh_i$ is smaller than the median of the
distribution $f_i$,
we have $P_i= 2P(\s_i < \sh_i)$, whereas  $\sh_i$ larger than the median
would give  $P_i= 2P(\s_i > \sh_i)$. The factor of two is present because
we want a two-sided test. Thus  $P_i = 1$ if $\sh_i$ equals the median,
$P_i = 2\%$ if $\sh_i$ is at the high 99th percentile, etc.

\subsection{2.2. Its probability distribution}

Apart from the simple interpretation of the
probability product $\eta$, it has the advantage that
its probability distribution can be calculated
analytically, and is completely independent of the
physics of the model --- in fact, it depends only
on $\N$. We will now give the exact distributions.

By construction, $0\leq\eta\leq 1$.
For $\N=1$, $\x$ will simply have a uniform distribution, {\ie}
$$\fx(\x) = \cases{
1&if $0\leq\x\leq 1$,\cr
0&otherwise.
}\eq$$
Thus in the general case, $\eta$ will be a product of $N$ independent
uniformly distributed random variables.
The calculation of the probability distribution for $\x$ is straighforward,
and can be found in a number of standard texts. The result is
$$\fx(\x) = -f_z(-\ln\x){dz\over d\x} =
\cases{
{1\over (\N-1)!}(-\ln\x)^{\N-1}&if $0\leq\x\leq 1$,\crr
0&otherwise.
}\eq$$

\subsection{2.3. The consistency probability}

The probability $P(\x<\xh)$, the
probability of getting as bad agreement as we do or worse,
is simply the cumulative distribution function $\Fx(\xh)$,
and the integral can be carried out analytically for any $\N$:
$$\Fx(\xh) \equiv P(\x<\xh) = \int_0^{\xh}\fx(u)du =
\xh\theta(\xh)\sum_{i=0}^{\N-1}{(-\ln\xh)^{\N}\over \N!},\eq$$
where $\theta$ is the Heaviside step function, and $\Fx(\xh)=1$ for
$\xh\geq 1$. Since the product of $\N$ numbers between zero and one tends
to zero as $\N\to\infty$, it is no surprise that
$$\Fx(\xh)\to\theta(\xh)\xh e^{-\ln\xh} = \theta(\xh)\eq$$
as $N\to\infty$, {\ie} that $\fx(\xh)\to\delta(\xh)$.
The function $\Fx(\xh)$ is plotted in Figure 1, and
the values of $\xh$ for which $\Fx(\xh) = 0.05$, 0.01 and 0.001,
respectively, are given in Table I for a few $N$-values.
For example, if three experimental results give a goodness-of-fit parameter
$\xh = 0.0002$ for some model, then this model is ruled out at a
confidence level of $99\%$.
Thus if the model where true and the experiments where repeated in very
many different horizon volumes of the universe, such a low
goodness-of-fit value would be obtained less than $1\%$ of the time.

\bigskip\goodbreak
\nobreak
{
\raggedright
\noindent
\baselineskip12pt
{\bf Table 1.} Probability product limits:
\tabskip = 1em
\halign{#\%&$#$\hfill&$#$\hfill&$#$\hfill&$#$\hfill\cr
\streck
\omit Confidence level&\N=1&\N=2&\N=3&\N=4\cr
\streck
95&0.05&0.0087&0.0018&0.00043\cr
99&0.01&0.0013&0.00022&0.000043\cr
99.9&0.001&0.000098&0.000013&0.0000021\cr
\streck
}}
\bigskip\goodbreak

\subsection{2.4. Ruling out whole classes of models}

If we wish to use the above formalism to test a whole set of models,
then we need to solve an optimization problem to find the one model
in the set for which the consistency probability is maximized.
For instance, if the family of models under consideration is standard $n=1$,
$\Gamma=0.5$ CDM (see Section 3), then the only free parameter is the overall
normalization constant $A$. Thus we can write the consistency probability as
$p(A)$, and use some numerical method to find the normalization $A_*$ for which
$p(A)$ is maximized.
After this, the
statistical interpretation is clear: if the experiments under consideration are
carried out in an ensemble of CDM universes, as extreme results as those
observed will only be obtained at most a fraction $p(A_*)$ of the time,
whatever the true normalization constant is. Precisely this case will be
treated in the next section.
For the slightly wider class of models
consisting of CDM power spectra with arbitrary $A$, $n$  and $\Gamma$, the
resulting optimization problem would be a three-dimensional one, and
the maximal consistency probability would necessarily satisfy
$$p(A_*,n_*,\Gamma_*) \ge p(A_*,1,0.5) = p(A_*).\eq$$
An even more general class of models is the set of all models where the
random fields are Gaussian, {\ie} allowing completely arbitrary power spectra
$P$. In section 4, we will show that the resulting infinite-dimensional
optimization problem can in be reduced to a succession of two
finite-dimensional
ones.

\beginsection{3. COLD DARK MATTER CONFRONTS SP91, COBE AND LAUER-POSTMAN}

As an example of an application of the formalism presented in the
previous section, we will now test the standard cold dark matter (CDM)
model of structure formation for consistency with the SP91 CMB
experiment and the Lauer-Postman bulk flow experiment.

Let $E_1$ be the Lauer-Postman (LP for short) measurement of bulk flows of
galaxies in a $150h^{-1}\Mpc$ sphere (Lauer and Postman, 1993).
Let $E_2$ be the 1991 South Pole CMB anisotropy experiment, SP91 for
short (Gaier {\etal} 1992).
Let $E_3$ be the COBE DMR experiment (Smoot {\etal} 1992).
All of these experiments probe scales that are well described by linear
perturbation theory, and so as long as the initial fluctuation are Gaussian,
the expected results of the experiments can
be expressed simply as integrals over the power spectrum of the matter
perturbation:
$$
\left<s_i\right>=\int W_i(k)P(k)dk.
$$
Here $s_{sp}$ and $s_{cobe}$ are the mean-square temperature fluctuations
measured by the experiments, and $s_{lp}\equiv (v/c)^2$ is the squared bulk
flow.
The corresponding
window functions $\Wlp$, $\Wsp$ and $\Wc$ are derived in Appendix A, and
plotted in Figure 2. These window functions assume  that the initial
perturbations were adiabatic, that $\Omega=1$, and that recombination
happened in the standard way,
{\i.e.} a last-scattering surface at $z\approx 1000$. The
SP91 window function is to be interpreted as a lower limit to the true window
function, as it includes contributions only from the Sachs-Wolfe effect, not
from Doppler motions or intrinsic density fluctuations of the
surface of last scattering.

Now let us turn to the probability distributions for the random variables
$\slp$, $\ssp$ and $\scobe$.
The standard
CDM model with power-law initial fluctuations
$\propto k^n$ predicts a power
spectrum that is well fitted by (Bond and Efstathiou 1984)
$$P(k) =
{A q^n\over
\left(1+\left[aq+(bq)^{1.5} +
(cq)^2\right]^{1.13}\right)^{2/1.13}},\eq$$
where $a\equiv 6.4$, $b\equiv 3.0$, $c\equiv 1.7$ and
$q\equiv (1h^{-1}\Mpc) k/\Gamma$.
For the simplest model, $\Gamma=h$, but certain additional complications
such as a non-zero cosmological constant $\Lambda$ and a non-zero fraction
$\Omega_{\nu}$ of hot dark matter can be fitted with reasonable accuracy
by other values of $\Gamma$ (Efstathiou, Bond and White 1992).
Thus the model has three free parameters: $n$, $\Gamma$ and the overall
normalization A. Integrating the power spectrum against
the three window functions yields the values of $c_i$ given in Table 2.
The two rightmost columns contain the quotients $\clp/\csp$ and $\clp/\ccobe$,
respectively.

\bigskip\goodbreak
\nobreak
{
\raggedright
\noindent
\baselineskip12pt
{\bf Table 2.} Expected r.m.s. signals for CDM power spectrum with
$A=(1h^{-1}\Mpc)^3$:
\tabskip = 1em
\halign{$#$&$#$\hfill&$#$\hfill&$#$\hfill&$#$&\hfill$#$&\hfill$#$\hfill\cr
\streck
n&\Gamma&\omit LP&\omit SP91&\omit COBE&\omit LP/SP91&\omit LP/COBE\cr
\streck
1&0.5&9.2\tt{-7}&1.6\tt{-8}&2.0\tt{-8}&56.7&45.1\cr
0.7&0.5&1.7\tt{-6}&2.9\tt{-8}&5.1\tt{-8}&57.2&32.7\cr
2&0.5&1.4\tt{-7}&2.6\tt{-9}&1.2\tt{-9}&53.9&112.2\cr
1&0.1&1.4\tt{-6}&2.3\tt{-8}&4.3\tt{-8}&57.9&31.9\cr
1&10&2.3\tt{-7}&4.1\tt{-9}&4.6\tt{-9}&55.9&49.6\cr
\streck
}}
\bigskip\goodbreak

\noindent
As can be seen, the dependence on $\Gamma$ is quite weak, and the quotient
$\clp/\csp$ is quite insensitive to the spectral index $n$ as well.
Let us for definiteness assume the canonical values $n=1$ and $\Gamma=0.5$
in what
follows.

These values $\c_i$ would be the average values of the probability
distributions for $\ssp$ and $\slp$ if there where no experimental
noise. We will now model the full probability distributions of the three
experiments, including the contribution from experimental noise.

For a bulk flow experiment, the
three components $v_x$, $v_y$ and $v_z$ of the velocity vector $\vv$ are
expected to be independent Gaussian random variables with zero mean, and
$$\expec{|\vv|^2} = \clp.\eq$$
However, this is not quite the random variable $\slp$ that we measure,
because of errors in distance estimation, {\it etc.} Denoting the
difference between the observed and true bulk velocity vectors by $\ve$,
let us assume that the three components of $\ve$ are identically
distributed and independent Gaussian random variables. This should be a
good approximation, since even if the errors for individual galaxies are
not, the errors in the average velocity $\ve$ will be approximately
Gaussian by the Central Limit Theorem. Thus the velocity vector that we
measure, $\vv+\ve$, is also Gaussian, being the sum of two Gaussians.
The variable that we actually measure is $\slp = |\vv+\ve|^2$, so
$$\slp = {1\over 3}\left(\clp + \vlp\right)\chi^2_3,\eq$$
where $\chi^2_3$
has a chi-squared distribution with three degrees of freedom,
and $\vlp$ is the variance due to experimental noise, {\ie} the
average variance that would be detected even if the true power spectrum
were $P(k)=0$.
The fact that the expectation value of the
detected signal $\slp$ (which is usually
referred to as the {\it uncorrected} signal in the literature)
exceeds the true signal $\clp$ is usually referred to as
{\it error bias}
(LP; Strauss Cen \& Ostriker 1993 -- hereafter SCO).
Error bias is ubiquitous to all
experiments of the type discussed in this paper, including
CMB experiments, since the
measured quantity is positive definite and the noise errors
contribute squared. In the
literature, experimentally detected signals are
usually quoted after error bias has been corrected for, {\ie}
after the noise has been subtracted from the uncorrected signal in
For LP, the uncorrected signal is 807 km/s, whereas the
signal quoted after error bias correction is 689 km/s.

For the special case of the LP experiment, detailed
probability distributions have been computed using Monte-Carlo
simulations
(LP, SCO),
which incorporate such experiment-specific complications as
sampling errors,  asymmetry in the error ellipsoid, etc.
To be used here, such simulations would need to be carried
out for each value of $\clp$ under consideration.
Since the purpose of this section is merely to
give an example of the test formalism, the above-mentioned
$\chi^2$-approximation will be quite sufficient for our needs.

For the SP91
nine-point scan, the nine true values $\Delta T_i/T$
are expected to be Gaussian random variables that to a good approximation
are independent. They have zero mean, and
$$\expec{\left|{\Delta T_i/T}\right|^2} = \csp.\eq$$
Denoting the difference
between the actual and observed values by $\delta_i$, we make the
standard assumption that these nine quantities are identically distributed
and independent Gaussian random variables.
Thus the temperature fluctuation that we measure at
each point,  $\Delta T_i/T+\delta$, is
again Gaussian, being the sum of two Gaussians.
The variable that we
actually measure is
$$\ssp =
{1\over 9} \sum_1^9 \left({\Delta T_i\over T}+\delta_i\right)^2
= {1\over 9}\left(\csp + \vsp\right)\chi^2_9,\eq$$
where $\chi^2_9$ has a chi-squared
distribution with nine degrees of freedom, and $\vsp$ is the variance
due to experimental noise, the error bias,
{\ie} the average variance that would be
detected even if the true power spectrum were $P(k)=0$.

We will use only the signal from highest of the four
frequency channels, which is the one likely to be the least affected
by galactic contamination. Again, although Monte-Carlo
simulations would be needed to obtain the exact probability
distributions, we will use the simple $\chi^2$-approximation here.
In this case, the main experiment-specific complication is the
reported gradient removal, which is a non-linear operation and thus
does not simply lead to a $\chi^2$-distribution with fewer degrees of
freedom.

The amplitude of the COBE signal can be characterized by the variance in
$\Delta T/T$ on an angular scale of $10^\circ$.  This number can be
estimated from the COBE data set as
$\scobe=\sigma^2_{10^\circ}= ((11.0\pm 1.8)\times
10^{-5})^2$ (Smoot {\etal} 1992).
The uncertainty in this quantity is purely due to
instrument noise, and contains no allowance for cosmic variance.  We
must fold in the contribution due to cosmic variance in order to determine
the probability distribution for $\scobe$.
We determined this probability distribution by performing Monte-Carlo
simulations of the COBE experiment.  We made simulated COBE maps with a
variety of power spectra (including power laws with indices ranging
from $0$ to $3$, as well as delta-function power spectra of the sort described
in Section 4).  We included instrumental noise in the maps, and excluded
all points within $20^\circ$ of the Galactic plane.  By estimating
$\scobe$ from each map, we were able to construct a probability
distribution corresponding to each power spectrum. In all cases, the
first three moments of the distribution were well approximated by
$$\eqalign{
\mu_1&\equiv\expec{\scobe}=\ccobe,\cr
\mu_2&\equiv\expec{\scobe^2}-
\expec{\scobe}^2\le 0.063 \ccobe^2+
1.44\times 10^{-21},\cr
\mu_3&\equiv\expec{\scobe^3}=0.009 \ccobe^3.\cr
}\eq$$
Furthermore, in all cases the probability distributions were well modeled
by chi-squared distributions with the number of degrees of freedom,
mean, and offset chosen to reproduce these three moments.  Note that
the magnitude of the cosmic variance depends on the shape of the power
spectrum as well as its amplitude.  The inequality in the above expression
for $\mu_2$ represents the largest cosmic variance of any of the power
spectra we tested.  Since we wish to set conservative limits on models,
we will henceforth assume that the cosmic variance is given by this
worst-case value.
Thus we are assuming that the random variable
$(\scobe-s_0)/\Delta s$ has a chi-squared distribution with
$\delta$ degrees of freedom, where
$$\eqalign{
\s_0&=\mu_1-2\mu_2^2/\mu_3,\cr
\Delta s&=\mu_3/4\mu_2,\cr
\delta&=8\mu_2^3/\mu_3^2.\cr
}\eq$$

The results obtained using these three probability distributions are
summarized in Tables 3abc. In 3a and 3b, $\N=2$, and the question
asked is whether LP is consistent with COBE and SP91, respectively. In
Table 3c, $\N=3$, and we test all three experiments for consistency
simultaneously. In each case, the optimum normalization (proportional to
the entries labeled ``Signal") is different, chosen such that
the consistency probability for the experiments under consideration is
maximized.
As can be seen, the last two tests rule out CDM at a significance
level of
$95\%$, {\ie}
predict that in an ensemble of universes, results as extreme
as those we observe would be obtained less than $5\%$ of the time.
Note that
using both COBE and SP91 to
constrain LP yields a rejection that is no stronger than that
obtained when ignoring COBE. In the latter case, the best fit is
indeed that with no cosmological power at all, which agrees well with
the observation of SCO
that sampling
variance would lead LP to detect a sizable bulk
flow (before correcting for error bias) even if there where none.

\bigskip\goodbreak
\nobreak
{
\raggedright
\noindent
\baselineskip12pt
{\bf Table 3a.} Are LP and COBE consistent with CDM?
\tabskip = 1em
\halign{#\hfill&$#$\hfill&$#$\hfill&$#$&\hfill$#$&\hfill$#$\hfill\cr
\streck
 &\omit LP&\omit COBE&\omit Combined\cr
\streck
Noise&420\kms&9.8\mK&\cr
Signal&169\kms&33.8\mK&\cr
Noise+Signal&453\kms&35.2\mK&\cr
Detected&807\kms&35.2\mK&\cr
$\xh$&0.046&1.00&0.046\cr
$P(\x<\xh)$&0.046&1.00&0.19\cr
\streck
}}
\bigskip\goodbreak

\bigskip\goodbreak
\nobreak
{
\raggedright
\noindent
\baselineskip12pt
{\bf Table 3b.} Are LP and SP91 consistent with CDM?
\tabskip = 1em
\halign{#\hfill&$#$\hfill&$#$\hfill&$#$&\hfill$#$&\hfill$#$\hfill\cr
\streck
 &\omit LP&\omit SP91&\omit Combined\cr
\streck
Noise&420\kms&26.4\mK&\cr
Signal&0\kms&0\mK&\cr
Noise+Signal&420\kms&26.4\mK&\cr
Detected&807\kms&19.9\mK&\cr
$\xh$&0.023&0.35&0.0079\cr
$P(\x<\xh)$&0.023&0.35&0.046\cr
\streck
}}
\bigskip\goodbreak

\bigskip\goodbreak
\nobreak
{
\raggedright
\noindent
\baselineskip12pt
{\bf Table 3c.} Are LP, SP91 and COBE all consistent with CDM?
\tabskip = 1em
\halign{#\hfill&$#$\hfill&$#$\hfill&$#$&\hfill$#$&\hfill$#$\hfill\cr
\streck
 &\omit LP&\omit SP91&\omit COBE&\omit Combined\cr
\streck
Noise&420\kms&26.4\mK&9.8\mK&\cr
Signal&168\kms&26.9\mK&33.8\mK&\cr
Noise+Signal&452\kms&37.7\mK&35.1\mK&\cr
Detected&807\kms&19.9\mK&35.2\mK&\cr
$\xh$&0.046&0.039&0.97&0.0017\cr
$P(\x<\xh)$&0.046&0.039&0.97&0.046\cr
\streck
}}
\bigskip

{\it
Tables 3abc show the consistency probability calculations.
The first line in each table gives the experimental noise, {\ie}
the detection that would be expected in the absence of any cosmological
signal.
The second line is the best-fit value for the cosmological signal $c$,
the value that maximizes the combined consistency probability
in the lower right corner of the table.
The third line contains
the expected value of an experimental detection, and
is the sum in quadrature of the two preceding lines.
The fourth line gives the goodness-of-fit parameter for
each of the experiments, i.e. the probability that they would yield
results at least as extreme as they did.
The rightmost number is the combined goodness-of-fit parameter, which is
the product
of the others.
The last line contains the consistency probabilities, the probabilities
of obtaining goodness-of-fit parameters at least as low as
those on the preceding line.
}
\vfill

\goodbreak

\beginsection{4. ALLOWING ARBITRARY POWER SPECTRA}

In this section, we will derive the mathematical formalism
for testing results from multiple experiments for consistency, without making
any assumptions whatsoever about the power spectrum.
This approach was pioneered by Juszkiewicz
{\etal} (1987) for the case $\N=2$. Here we generalize the
results to the case of arbitrary $\N$.
Despite the fact that the original optimization problem is
infinite-dimensional, the necessary calculations will be seen to be of a
numerically straightforward type, the case of $\N$ independent constraints
leading
to nothing more involved than numerically solving a system of $n$ coupled
non-linear equations. After showing this, we will discuss some inequalities
that provide  both a good approximation of the exact results and a
useful qualitative understanding of them.

\subsection{4.1. The Optimization Problem}

Let us consider $N = n+1$ experiments numbered 0, 1, ..., $n$ that
probe the cosmological power spectrum $P(k)$. We will think of each
experiment as measuring some weighted average of the power spectrum,
and characterize an experiment $E_i$ by its window function
$W_i(k)$ as before.

Purely hypothetically, suppose we that we had repeated the same
experiments in many different locations in the universe, and for all
practical purposes knew the quantities $c_1,...,c_n$ exactly. Then for
which power spectrum $P(k)$ would $c_0$ be maximized, and what would
this maximum be? If we experimentally determined $c_0$ to be larger
than this maximum value, our results would be inconsistent, and we
would be forced to conclude that something was fundamentally wrong
either with our theory or with one of the experiments. In this section, we will
solve this hypothetical problem. After this, it will be seen that the real
problem, including cosmic variance and experimental noise, can be solved in
almost exactly the same way.

The extremal power spectrum we are looking for is the solution to the
following linear variational problem:

Maximize
\ContMinEq=\eqnr
$$\intk P(k) W_0(k) dk\eq$$
subject to the constraints that
$$\cases{
\intk P(k) W_i(k) dk = c_i&for $=1,...,n$,\crr
P(k) \geq 0&for all $k\geq 0.$
}\eq$$

This is the infinite-dimensional analogue of the so called
linear programming problem, and its solution is quite analogous to
the finite-dimensional case.
In geometrical terms, we think of each power spectrum as a point in the
infinite dimensional vector space of power spectra (tempered
distributions on the positive real line, to be precise), and limit
ourselves to the subset $\Omega$ of points where all the above
constraints are satisfied. We have a linear function on this space,
and we seek the point within the subset $\Omega$ where this function is
maximized.
We know that a differentiable functional on a bounded region takes its
maximum either at an interior point, at which its gradient will
vanish, or at a boundary point.
In linear optimization problems like the one above, the
gradient (here the variation with respect to $P$, which is simply
the function $W_0$) is simply a constant, and will never vanish.
Thus any maximum will always be attained at a boundary point.
Moreover, from the theory of linear programming, we know that if
there are $n$ linear constraint equations, then the optimum point
will be a point where all but at most $n$ of the coordinates are zero.
It is straightforward to generalize this result to our
infinite-dimensional case, where each fixed $k$ specifies
a ``coordinate" $P(k)$, and the result is that the solution to the
variational problem is of the form
$$P(k) = \sumi p_i \delta(k-k_i).\eq$$
This reduces the optimization problem from an infinite-dimensional
one to a $2n$-dimensional one, where only the constants $p_i$ and $k_i$
remain to be determined:

Maximize
\DiscMinEq=\eqnr
$$\sumj p_j W_0(k_j)\eq$$
subject to the constraints that

$$\cases{
\sumj p_j W_i(k_j) = c_i&for $i=1,...,n$,\crr
p_i \geq 0&for $i=1,...,n$.
}\eq$$
This problem is readily solved using the method of Lagrange multipliers:
defining the Lagrangian
$$L = \sumj p_j W_0(k_j) -
\sumi\lambda_i\left[\sumj p_j W_i(k_j) - c_i\right]\eq$$
and requiring that all derivatives vanish leaves the following set of $3n$
equations to determine the $3n$ unknowns $p_i$, $k_i$ and
$\lambda_i$:
$$\left\{
\eqalign{
W_0(k_i) - \sumj \lambda_j W_j(k_i)&= 0,\crr
\left[W'_0(k_i) - \sumj \lambda_j W'_j(k_i)\right]p_i&= 0,\crr
c_i - \sumj p_j W_i(k_j)&= 0.
}
\right.\eq$$
Introducing matrix notation by defining the $k_i$-dependent quantities
$A_{ij}\equiv W_j(k_i)$, $B_{ij}\equiv W'_j(k_i)$,
$a_i\equiv W_0(k_i)$ and $b_i\equiv W'_0(k_i)$ brings out the
structure of these equations more clearly:
If $p_i\neq 0$, then
$$\left\{
\eqalign{
A\vl&= \va,\crr
B\vl&= \vb,\crr
A^T\vp &= \vc.
}
\right.\eq$$
If $A$ and $B$ are invertible, then eliminating $\vl$ from the
first two equations yields the following system of $n$
equations to be solved for the $n$
unknowns $k_1$, ..., $k_n$:
\nonlinEq=\eqnr
$$A^{-1}\va = B^{-1}\vb.\eq$$
Although this system is typically coupled and non-linear and out of
reach of analytical solutions for realistic window functions, solving
it numerically is quite straightforward.
A useful feature is that once this system is solved, $\va$, $\vb$, $A$ and
$B$ are mere constants, and the other unknowns are simply given by
matrix inversion: $$\left\{
\eqalign{
\vl&= A^{-1}\va\crr
\vp&= \left(A^{-1}\right)^T\vc\crr
}
\right.\eq$$
Since the non-linear system $(\number\nonlinEq)$ may have more than
one solution, all solutions should be substituted back into
$(\number\DiscMinEq)$ to determine which one is the
global minimum.
Furthermore, to make statements about the solution to our original
optimization problem $(\number\ContMinEq)$, we need to consider also
the case where one or more of the $n$ variables $p_1$,...,$p_n$ vanish.
If exactly $m$ of them are non-vanishing, then without loss of
generality, we may assume that these are the first $m$ of the $n$
variables. Thus we need to solve the maximization problem
$(\number\DiscMinEq)$ separately for the cases where $P(k)$ is
composed of $n$ delta functions, $n-1$ delta functions, {\it etc.},
all the way down to the case where $P(k)$ is single delta function.
These solutions should then be substituted back into
$(\number\ContMinEq)$ to determine which is the global minimum sought
in our original problem.
Thus the solutions depend on the window functions $W_i$ and the
signals $c_i$ in the following way:

\blist{From the window functions alone, we can determine a discrete
and usually finite number of candidate wavenumbers $k$ where delta
functions can be placed.}

\blist{The actual signals $c_i$ enter only in determining the
coefficients of the delta functions in the sum, i.e. in determining
what amount of power should be hidden at the various candidate
wavenumbers.}

\noindent
If we have found an the optimal solution, then a small change in the
signal vector $\vc$ will typically result in a small change in $\vp$
and no change at all in the number of delta functions in $P(k)$ or
their location. If $\vc$ is changed by a large enough amount, the
delta functions may suddenly jump and/or change in number as a
different solution of $(\number\DiscMinEq)$ takes over as global
optimum or one of the coefficients $p_i$ becomes negative, the latter
causing the local optimum to be rejected for constraint violation.
Thus within certain limits, we get the extremely simple result that
for the optimal power spectrum $P(k)$,
$$c_0 = \intk P(k) W_0(k) dk =
\left(A^{-1}\va\right)\cdot\vc.\eq$$
Thus within these limits,
$c_0$ depends linearly on the observed
signal strengths $c_i$. This is exactly analogous to what happens in
linear programming problems.

\subsection{4.2. A Useful Inequality}

Before proceeding further, we will attempt to provide a more
intuitive understanding of the results of the previous section, and show how to
determine how complicated a calculation is justified.
For the special case of only
a single constraint, {\ie} $n=1$, we obtain simply $P(k) = p_1\delta(k-k_1)$,
where $k_1$ is given by
$$W_0'(k_1)W_1(k_1) = W_1'(k_1)W_0(k_1).\eq$$
For the case of $n$ constraints, let us define the functions
$$\f_i\equiv  {W_0(k)\over W_1(k)} c_i.$$
Then we see that for $n=1$, $k_1$ is simply the wavenumber for which the
function $\f_1$ is maximized, and that the maximum signal possible is simply
$c_0 = \f_1(k_1)$.
Thus the maximum signal in experiment $0$ that is consistent with
the constraint from experiment $i$ is obtained when
the power is concentrated where the function $\f_i$ is
large. In other words, if we want to explain a high signal $c_0$ in the face
of low signals in several constraining experiments, then the best
place to hide the necessary power from the $i^{th}$ experiment is where $f_i$
takes its maximum.
These functions are plotted in Figure 4 for the experiments discussed in
Section 3, the optimization problem being the search for
the maximum LP signal that is consistent with the constraints from SP91
and COBE. For illustrative purposes, we here assume that $\csp$ and $\ccobe$
are
known exactly, and given by the detected signals
$\ssph^{1/2}\approx 19.9\mK$ and $\scobeh^{1/2}\approx 33.8\mK$
(we will give a
proper treatment of cosmic variance and noise in the following section).
Using the $n=1$ constraint for each constraining experiment separately, the
smallest of the functions thus sets an
upper limit to the allowed signal $c_0=\csp$.
Thus the limit is given by the highest point in the hatched region in Figure
4, {\ie}
$$c_0\leq c_{max}^{(1)}\equiv\sup_k\>\min_i f_i(k).\eq$$
We see that using the SP91 constraint
alone, the LP signal would be maximized if all power were at $k\approx
(940\Mpc)^{-1}$. Since this flagrantly violates the COBE constraint, the best
place to hide the power is instead at $k\approx (100\Mpc)^{-1}$.

By using the above formalism to impose all the constraints at once, the
allowed signal obviously becomes lower. If the constraints are equalities
rather
than inequalities, then this stronger limit can
never lie below value at $(k_*\approx 250\Mpc)^{-1}$, where
$\fsp(k_*)=\fcobe(k_*)$, since this is the signal that would result from a
power
spectrum of the form $P(k)\propto\delta(k-k_*)$. Thus for the particular window
functions in our example, where the constraint from the $n=2$ calculation
cannot
be more than a factor $\fsp(80\Mpc)/\fsp(250\Mpc)\approx 1.05$ stronger than
the simple $n=1$ limits, the latter are so close to the true optimum that they
are quite sufficient for our purposes.
If the constraints are upper limits
rather than equalities, then the limit on $c_0$ is more relaxed, and is always
the uppermost point in the hatched region, {\ie} $c_{max}^{(1)}$.

\subsection{4.3. Including Noise and Cosmic Variance}

To correctly handle cosmic variance and instrumental noise, we need to use the
formalism developed in Section 2.
Thus given the probability distributions for the various experimental results
$s_i$, we wish to find the power spectrum for which the consistency probability
$\x$ is maximized.
This optimization problem, in which
all experiments are treated on an equal footing, will be seen to lead directly
to
the asymmetric case above where the signal in one is
maximized given constraints from the others.
For definiteness, we will continue using the example with the LP, SP91
and COBE experiments.
As seen in Section 2, the source of the low consistency probabilities is that
$\slph$ is quite high when compared to $\ssph$ and $\scobeh$. Thus
it is fairly obvious that for the power spectrum that maximizes the
consistency probability, we will have $\slph > \expec{\slp}$, whereas
$\ssph < \expec{\ssp}$ and $\scobeh < \expec{\scobe}$, so we can neglect power
spectra that do not have this property. Let us first restrict ourselves to
the subset of these power spectra for which  $\clp = D$ and $\ccobe = E$, where
$D$ and $E$ are some constants.
Then these power spectra all predict the same probability distributions for
$\slp$ and $\scobe$. The consistency probability $\eta$ is clearly maximized
by the power spectrum that maximizes $\expec{\ssp}$, and this will be a
linear combination of one or two delta functions as shown in Section 4.1. The
key point is that since the locations of these delta functions are
independent of $D$ and $E$ (within the range discussed in 4.1), the
infinite-dimensional optimization problem reduces to the following two
simple steps:

\llist{1. Solve for the optimal number of delta functions $m$ and their
locations
$k_i$ as described in section 4.1}

\llist{2. Find the $m$ coefficients $p_i$ for which the power spectrum
$P(k) = \sum_{i=1}^m p_i\delta(k-k_i)$ maximizes the consistency probability.}

\subsection{4.4. Power spectrum independent constraints on LP, SP91 and COBE}

When applying the above consistency test to the LP, SP91 and COBE
experiments, we obtain exactly the same consistency probability
as in Table 3b. The reason for this is that the optimal
normalization turns out to be zero. This will obviously change if the
LP error bars become smaller in the future.
Thus dropping the CDM assumption does not improve the situation
at all, which indicates that main source of the
inconsistency must be something other than the CDM model.

In anticipation of future developments,
consistency probabilities where also computed for a number of cases
with less noise in the LP experiment.
Comparing only LP and SP91, the optimum power spectrum has a delta function at
$k\approx (941\Mpc)^{-1}$.
When including all three experiments, treating the COBE and SP91
constraints as upper limits,
the optimum power spectrum has a single delta
function at $k\approx (79\Mpc)^{-1}$, so
the addition of COBE strengthens the constraint only
slightly, due to the
flatness of $\fsp$ in Figure 4.
Interestingly, for all these cases with smaller LP error bars,
consistency  probabilities were found to be almost as low when
allowing arbitrary power spectra as for the CDM case. This is again
attributable to the flatness of $\fsp$, since weighted averages of a
flat function are fairly independent of the shape of the weight
function (here the power spectrum).

\beginsection{5. DISCUSSION}

We have developed a formalism for testing multiple cosmological experiments for
consistency.
As an example of an application, we have used it to place
constraints on bulk flows of
galaxies using the COBE and SP91 measurements of fluctuations in the cosmic
microwave background.
It was found that taken at face value,
the recent detection by Lauer and Postman of a bulk
flow of $689$ km/s on scales of $150h^{-1}$Mpc
is inconsistent with SP91
within the framework of a CDM
model, at a significance level of about 95\%.
However, interestingly, this cannot be due solely to the CDM assumption,
since the LP result was shown to be inconsistent with COBE and
SP91 at the same significance level even when no assumptions whatsoever
were made about the power
spectrum. This leaves four possibilities:

\llist{
1. The window functions are not accurate.}

\llist{
2. Something is wrong with the quoted signals
or error bars for at least one of the experiments,
}

\llist{
3. The observed fluctuations cannot be explained within the framework
of gravitational instability and the Sachs-Wolfe effect.
}

\llist{
4. The random fields are not Gaussian,
}

\noindent
Case 1 could be attributed to a number of effects:
If $\Omega\neq 1$, then both the calculation of the Sachs-Wolfe
effect (which determines $\Wsp$ and $\Wc$) and the growth of
velocity perturbations (which determines $\Wlp$) are altered.
If the universe became reionized early enough to rescatter a significant
fraction of all CMB photons, then small scale CMB anisotropies
were suppressed, which would lower $\Wsp$. A quantitative treatment of these
two cases will be given in a future paper.
Other possible causes of 1 include
a significant fraction of the density
perturbations being isocurvature (entropy) perturbations or
tensor-mode perturbations
(gravity waves).
Apart from these uncertainties, we have made several
simplifying assumptions about the window functions for LP and SP91.
To obtain more accurate consistency probabilities than those derived
in the present paper, a more accurate LP window function should be used
that incorporates the discreteness and the asymmetry of
the sample of Abell clusters used. This can either be done
analytically (Feldman \& Watkins 1993) or circumvented altogether
by performing Monte-Carlo simulations like those of LP or
SCO, but for the whole family of power spectra under consideration.

As to case 2, there has been considerable debate about both the LP and
the SP91 experiments.
A recent Monte-Carlo Simulation of LP by SCO
basically confirms the large error bars quoted by LP. As is evident from the
flatness of LP curve in Figure 3, it will be impossible to make very strong
statements about inconsistency until future experiments produce smaller error
bars.
With the SP91 experiment, a source of concern is the validity of using
only the highest of the four frequency channels to place limits, even though
it is fairly clear that the other three channels suffer from problems with
galactic contamination. The situation is made more disturbing by the fact that
a
measurement by the balloon-borne MAX experiment (Gundersen {\etal} 1993) has
produced detections of degree-scale fluctuations  that that are higher than
those seen by SP91, and also higher than another MAX measurement (Meinhold
{\etal} 1993). On the other hand, SP91 has been used only as an upper limit
in our treatment, by including only the Sachs-Wolfe effect
and neglecting both
Doppler contributions
from peculiar motions of the surface of last scattering
and intrinsic density fluctuations at the recombination epoch.
If these effects (which unfortunately depend strongly on parameters such as
$h$ and $\Omega_b$) where included, the resulting constraints
would be stronger.

Case 3 might be expected if the universe
underwent a late-time
phase transition, since this could generate new large-scale
fluctuations in an entirely non-gravitational manner.

In the light of the many caveats in categories 1 and 2,
the apparent
inconsistency between LP and SP91 (Jaffe {\etal} 1993)
is hardly a source of major concern at the present time,
and it does not appear necessary to invoke 3 or 4.
However, we expect the testing formalism developed in this paper
to be able to provide many useful constraints in the future, as
more experimental data is accumulated and error bars become smaller.

\bigskip
The authors wish to thank Michael Strauss,
Bhuvnesh Jain, Douglas Scott, Martin White,
and Joseph Silk for many useful comments.

\baselineskip12pt
\frenchspacing

\beginsection REFERENCES

% \narrower
\smallskip

\rf Alcock, C, Akerlof, C. W., Allsman, R. A., Axelrod, T. S. and others 1993;
Nature;365;621

\rf Bond, J.R. \& Efstathiou, G. 1987;MNRAS;226;655

\rf Bond, J.R., Efstathiou, G., Lubin, P.M., \& Meinhold, P.R. 1991;
Phys. Rev. Lett.;45;1980

\rn
Cen, R., Gnedin, N. Y., Koffmann, L. A., \& Ostriker, J. P.
1992, preprint.

\rf Cen, R., Ostriker, J. P. \& Peebles, P. J. E. 1993;Ap. J.;415;423

\rf Dodelson, S. \& Jubas, J.M. 1993; Phys. Rev. Lett.;70;2224

\rn Feldman, H. A. \& Watkins, R. 1993, preprint.

\rf Feynman, R. P. 1939;Phys. Rev.;56;340

\rf Gaier, T., Schuster, J., Gundersen, J. O., Koch, T., Meinhold, P. R.,
Seiffert, M. \& Lubin, P. M., 1992;ApJ;398;L1

\rf Gnedin, N. Y. \& Ostriker, J. P.1992;Ap. J.;400;1

\rf G\'orski, K.1991;Ap. J. Lett.;370;L5

\rf G\'orski, K.1992;Ap. J. Lett.;398;L5

\rf Gundersen, J.O., Clapp, A.C., Devlin, M., Holmes, W., Fischer, M.L.,
Meinhold, P.R., Lange, A.E., Lubin, P. M., Richards, P.L., \& Smoot, G.F.
1993;Ap. J. Lett.;413;L1

\rn Jaffe, A., Stebbins, A., \& Frieman, J.A 1993.  Ap. J. (in press)

\rn Juszkiewicz, R. 1993, Private communication.

\rf Juszkiewicz, R., G\'orski, K., \& Silk, J.1987;Ap. J. Lett.;323;L1

\rn Kamionowski, M. \& Spergel, D. 1993, preprint.

\rn Kolb, E. W. \& Turner, M.S. 1990, {\it The Early Universe}, Addison
Wesley

\rn Lauer, T. \& Postman, M. 1993, preprint. (LP)

\rf Maddox, S. J., Efstathiou, G., Sutherland, W. J.
\& Loveday, J. 1990;MNRAS;242;43

\rf Meinhold, P. R. \& Lubin, P. M. 1991;Ap. J. Lett.;370;L11

\rf Meinhold, P. R., Clapp, A.C., Cottingham, D., Devlin, M., Fischer, M.L.,
Gundersen, J.O., Holmes, W., Lange, A.E., Lubin, P.M., Richards, P.L., \&
Smoot, G.F. 1993;Ap. J. Lett.;409;L1

\rf Oukbir, J. \& Blanchard, A. 1992;Astr. Ap.;262;L21

\rn Padmanabhan, T. 1993, {\it Structure Formation in the Universe},
Cambridge Univ. Press, New York

\rf Peebles, P. J. E. 1984; Ap. J.;284;439

\rf Peebles, P. J. E. 1987; Ap. J.;315;L73

\rn Schlegel, D., Davis, M., Summers, F. \& Holtzman, J. 1993, preprint.

\rf Smoot, G. F. {\etal} 1992; ApJ;396;L1

\rn Strauss, M., Cen, R. \& Ostriker, J.P. 1993, preprint. (SCO)

\rf Suto, Y., G\'orski, K. Juszkiewicz, R., \& Silk, J.1988; Nature;332;328

\rf Watson, R.A., Guti\'errez de la Cruz, C.M., Davies, R.D., Lasenby, A.N.,
Rebolo, R., Beckman, J.E., \& Hancock, S.1992;Nature;357;660

\rn Watson, R.A. \& Guti\'errez de la Cruz, C.M. 1993, preprint.

\rn White, M., Krauss, L., \& Silk, J. 1993, preprint.

\vfill\eject

\beginsection{APPENDIX: WINDOW FUNCTIONS}

\def\rhat{\hat{\bf r}}
The results of CMB anisotropy experiments can be conveniently described
by expanding the temperature fluctuation in spherical harmonics:
$$
{\Delta T\over T}(\rhat)=\sum_{l=2}^\infty\sum_{m=-l}^l a_{lm}Y_{lm}(\rhat).
\eq$$
(The monopole and dipole anisotropies
have been removed from the above expression, since
they are unmeasurable.)
If the fluctuations are Gaussian, then each coefficient $a_{lm}$ is an
independent Gaussian random variable with zero mean (Bond and Efstathiou
1987).  The statistical
properties of the fluctuations are then completely specified by the
variances of these quantities
$$
C_l\equiv \left<|a_{lm}|^2\right>.
\eq$$
(The fact that the variances are independent of $m$ is an immediate
consequence of spherical symmetry.)  Different CMB experiments are sensitive
to different linear combinations of the $C_l$'s:
\TedEqOne=\eqnr
$$
S=\sum_{l=2}^\infty F_lC_l,\eq
$$
where $S$ is the ensemble-averaged mean-square signal in a particular
experiment, and the ``filter function'' $F_l$ specifies the sensitivity
of the experiment on different angular scales.  The filter functions for
COBE and SP91 are
\TedEqTwo=\eqnr
$$\eqalign{
F_l^{(cobe)}&={(2l+1)\over 4\pi}e^{-\sigma_c^2(l+{1\over 2})^2},\cr
F_l^{(sp)}&=4e^{-\sigma_s^2(l+{1\over 2})^2}\sum_{m=-l}^l H_0^2(\alpha m),
}\eq$$
where $H_0$ is a Struve function.
$\sigma_c=4.25^\circ$ and $\sigma_s=0.70^\circ$ are the {r.m.s.} beamwidths
for the two experiments, and $\alpha=1.5^\circ$ is the amplitude of the
beam chop (Bond {\etal} 1991, Dodelson and Jubas 1993, White {\etal} 1992).

For Sachs-Wolfe fluctuations in a spatially flat Universe with the standard
ionization history, the angular power spectrum $C_l$ is related to the power
spectrum of the matter fluctuations in the following way
(Peebles 1984, Bond and Efstathiou 1987):
\TedEqThree=\eqnr
$$
C_l={8\over \pi\tau_0^4} \int_0^\infty dk P(k)\jbar_l^2(k).\eq
$$
Here $\tau_0$ is the conformal time at the present epoch, and
$$
\jbar_l(k)\equiv\int j_l(k\tau)V(k\tau)\,d\tau,\eq
$$
where $j_l$ is a spherical Bessel function.
The visibility function $V$ is the probability distribution for the conformal
time at which a random CMB photon was last scattered.  $\jbar_l(k)$
is therefore the average of $j_l(k\tau)$ over the last scattering surface.
We have used the $V$ of Padmanabhan (1993).

We can combine equations
$(\number\TedEqOne)$,
$(\number\TedEqTwo)$, and
$(\number\TedEqThree)$
to get the window
functions for the two experiments:
$$\eqalign{
\Wc &= {2\over\pi^2k^2\tau_0^4}\sum_{l=2}^{\infty}
\jbar_l^2(k)e^{-\sigma_c^2\left(l+{1\over 2}\right)^2}(2l+1)\cr
\Wsp &= {32\over\pi k^2\tau_0^4}\sum_{l=2}^{\infty}
\jbar_l^2(k)
e^{-\sigma_s^2\left(l+{1\over 2}\right)^2}
\sum_{m=-l}^l H_0^2(\alpha m)\cr
}\eq$$

The mean-square bulk flow inside of a sphere of radius $a$ is
(see, {\it e.g.}, Kolb and Turner 1990)
\TedEqFour=\eqnr
$$
\left<v^2\right>=\int dk P(k){18\over \pi^2\tau_0^2}\, {j_1^2(ka)\over (ka)^2}.
\eq
$$
However, we must make two corrections to this result before applying it
to the LP data.  This formula applies to a measurement of the bulk
flow within a sphere with an infinitely sharp boundary.  In reality,
errors in measuring distances cause the boundary of the spherical region
to be somewhat fuzzy.  If we assume that distance measurements are subject
to a fractional error $\epsilon$, then the window function must be
multiplied by $e^{-(\epsilon ka)^2}$.  We have taken $\epsilon=0.16$,
the average value quoted by LP. It should be noted that this value varies
from galaxy to galaxy in the LP sample, due to the distance estimation
technique used, and that a  more accurate window function that reflects
the discrete locations of the Abell clusters used in the survey
should take this into account.

The second correction has to do with the behavior of the window function
at small $k$.  Equation $(\number\TedEqFour)$ applies to the velocity
relative to the rest frame of the Universe.  The velocity measured by LP is
with respect to the CMB rest frame.  If there is an intrinsic CMB dipole
anisotropy, then these two reference frames differ.  Therefore, we must
include in equation $(\number\TedEqOne)$,
a term corresponding to the intrinsic CMB dipole.  This correction was first
noticed by G\'orski (1991).
After applying both of these corrections, the LP window function is
$$
W_{lp}={18\over \pi^2\tau_0^2}\left({j_1(ka)\over ka}e^{-(\epsilon ka)^2}
-{\jbar_1(k\tau_0)\over k\tau_0}\right)^2.
\eq
$$

\end  % HELLO! TO PRINT THE FIGURES AS WELL, SIMPLY DELETE THIS LINE.

\ifx\undefined\psfig\else\endinput\fi

%
% from a suggestion by eijkhout@csrd.uiuc.edu to allow
% loading as a style file:
\edef\psfigRestoreAt{\catcode`@=\number\catcode`@\relax}
\catcode`\@=11\relax
\newwrite\@unused
\def\ps@typeout#1{{\let\protect\string\immediate\write\@unused{#1}}}
\ps@typeout{psfig/tex 1.8}

%% Here's how you define your figure path.  Should be set up with null
%% default and a user useable definition.

\def\figurepath{./}

%
% @psdo control structure -- similar to Latex @for.
% I redefined these with different names so that psfig can
% be used with TeX as well as LaTeX, and so that it will not
% be vunerable to future changes in LaTeX's internal
% control structure,
%
\def\@nnil{\@nil}
\def\@empty{}
\def\@psdonoop#1\@@#2#3{}
\def\@psdo#1:=#2\do#3{\edef\@psdotmp{#2}\ifx\@psdotmp\@empty \else
    \expandafter\@psdoloop#2,\@nil,\@nil\@@#1{#3}\fi}
\def\@psdoloop#1,#2,#3\@@#4#5{\def#4{#1}\ifx #4\@nnil \else
       #5\def#4{#2}\ifx #4\@nnil \else#5\@ipsdoloop #3\@@#4{#5}\fi\fi}
\def\@ipsdoloop#1,#2\@@#3#4{\def#3{#1}\ifx #3\@nnil
       \let\@nextwhile=\@psdonoop \else
      #4\relax\let\@nextwhile=\@ipsdoloop\fi\@nextwhile#2\@@#3{#4}}
\def\@tpsdo#1:=#2\do#3{\xdef\@psdotmp{#2}\ifx\@psdotmp\@empty \else
    \@tpsdoloop#2\@nil\@nil\@@#1{#3}\fi}
\def\@tpsdoloop#1#2\@@#3#4{\def#3{#1}\ifx #3\@nnil
       \let\@nextwhile=\@psdonoop \else
      #4\relax\let\@nextwhile=\@tpsdoloop\fi\@nextwhile#2\@@#3{#4}}
%
% \fbox is defined in latex.tex; so if \fbox is undefined, assume that
% we are not in LaTeX.
% Perhaps this could be done better???
\ifx\undefined\fbox
% \fbox code from modified slightly from LaTeX
\newdimen\fboxrule
\newdimen\fboxsep
\newdimen\ps@tempdima
\newbox\ps@tempboxa
\fboxsep = 3pt
\fboxrule = .4pt
\long\def\fbox#1{\leavevmode\setbox\ps@tempboxa\hbox{#1}\ps@tempdima\fboxrule
    \advance\ps@tempdima \fboxsep \advance\ps@tempdima \dp\ps@tempboxa
   \hbox{\lower \ps@tempdima\hbox
  {\vbox{\hrule height \fboxrule
          \hbox{\vrule width \fboxrule \hskip\fboxsep
          \vbox{\vskip\fboxsep \box\ps@tempboxa\vskip\fboxsep}\hskip
                 \fboxsep\vrule width \fboxrule}
                 \hrule height \fboxrule}}}}
\fi
%
%%%%%%%%%%%%%%%%%%%%%%%%%%%%%%%%%%%%%%%%%%%%%%%%%%%%%%%%%%%%%%%%%%%
% file reading stuff from epsf.tex
%   EPSF.TEX macro file:
%   Written by Tomas Rokicki of Radical Eye Software, 29 Mar 1989.
%   Revised by Don Knuth, 3 Jan 1990.
%   Revised by Tomas Rokicki to accept bounding boxes with no
%      space after the colon, 18 Jul 1990.
%   Portions modified/removed for use in PSFIG package by
%      J. Daniel Smith, 9 October 1990.
%
\newread\ps@stream
\newif\ifnot@eof       % continue looking for the bounding box?
\newif\if@noisy        % report what you're making?
\newif\if@atend        % %%BoundingBox: has (at end) specification
\newif\if@psfile       % does this look like a PostScript file?
%
% PostScript files should start with `%!'
%
{\catcode`\%=12\global\gdef\epsf@start{%!}}
\def\epsf@PS{PS}
\def\epsf@getbb#1{%
%
%   The first thing we need to do is to open the
%   PostScript file, if possible.
%
\openin\ps@stream=#1
\ifeof\ps@stream\ps@typeout{Error, File #1 not found}\else
%
%   Okay, we got it. Now we'll scan lines until we find one that doesn't
%   start with %. We're looking for the bounding box comment.
%
   {\not@eoftrue \chardef\other=12
    \def\do##1{\catcode`##1=\other}\dospecials \catcode`\ =10
    \loop
       \if@psfile
	  \read\ps@stream to \epsf@fileline
       \else{
	  \obeyspaces
          \read\ps@stream to \epsf@tmp\global\let\epsf@fileline\epsf@tmp}
       \fi
       \ifeof\ps@stream\not@eoffalse\else
%
%   Check the first line for `%!'.  Issue a warning message if its not
%   there, since the file might not be a PostScript file.
%
       \if@psfile\else
       \expandafter\epsf@test\epsf@fileline:. \\%
       \fi
%
%   We check to see if the first character is a % sign;
%   if so, we look further and stop only if the line begins with
%   `%%BoundingBox:' and the `(atend)' specification was not found.
%   That is, the only way to stop is when the end of file is reached,
%   or a `%%BoundingBox: llx lly urx ury' line is found.
%
          \expandafter\epsf@aux\epsf@fileline:. \\%
       \fi
   \ifnot@eof\repeat
   }\closein\ps@stream\fi}%
%
% This tests if the file we are reading looks like a PostScript file.
%
\long\def\epsf@test#1#2#3:#4\\{\def\epsf@testit{#1#2}
			\ifx\epsf@testit\epsf@start\else
\ps@typeout{Warning! File does not start with `\epsf@start'.  It may not be a
PostScript file.}
			\fi
			\@psfiletrue} % don't test after 1st line
%
%   We still need to define the tricky \epsf@aux macro. This requires
%   a couple of magic constants for comparison purposes.
%
{\catcode`\%=12\global\let\epsf@percent=%\global\def\epsf@bblit{%BoundingBox}}
%
%
%   So we're ready to check for `%BoundingBox:' and to grab the
%   values if they are found.  We continue searching if `(at end)'
%   was found after the `%BoundingBox:'.
%
\long\def\epsf@aux#1#2:#3\\{\ifx#1\epsf@percent
   \def\epsf@testit{#2}\ifx\epsf@testit\epsf@bblit
	\@atendfalse
        \epsf@atend #3 . \\%
	\if@atend
	   \if@verbose{
		\ps@typeout{psfig: found `(atend)'; continuing search}
	   }\fi
        \else
        \epsf@grab #3 . . . \\%
        \not@eoffalse
        \global\no@bbfalse
        \fi
   \fi\fi}%
%
%   Here we grab the values and stuff them in the appropriate definitions.
%
\def\epsf@grab #1 #2 #3 #4 #5\\{%
   \global\def\epsf@llx{#1}\ifx\epsf@llx\empty
      \epsf@grab #2 #3 #4 #5 .\\\else
   \global\def\epsf@lly{#2}%
   \global\def\epsf@urx{#3}\global\def\epsf@ury{#4}\fi}%
%
% Determine if the stuff following the %%BoundingBox is `(atend)'
% J. Daniel Smith.  Copied from \epsf@grab above.
%
\def\epsf@atendlit{(atend)}
\def\epsf@atend #1 #2 #3\\{%
   \def\epsf@tmp{#1}\ifx\epsf@tmp\empty
      \epsf@atend #2 #3 .\\\else
   \ifx\epsf@tmp\epsf@atendlit\@atendtrue\fi\fi}

% End of file reading stuff from epsf.tex
%%%%%%%%%%%%%%%%%%%%%%%%%%%%%%%%%%%%%%%%%%%%%%%%%%%%%%%%%%%%%%%%%%%

%%%%%%%%%%%%%%%%%%%%%%%%%%%%%%%%%%%%%%%%%%%%%%%%%%%%%%%%%%%%%%%%%%%
% trigonometry stuff from "trig.tex"
\chardef\letter = 11
\chardef\other = 12

\newif \ifdebug %%% turn me on to see TeX hard at work ...
\newif\ifc@mpute %%% don't need to compute some values
\c@mputetrue % but assume that we do

\let\then = \relax
\def\r@dian{pt }
\let\r@dians = \r@dian
\let\dimensionless@nit = \r@dian
\let\dimensionless@nits = \dimensionless@nit
\def\internal@nit{sp }
\let\internal@nits = \internal@nit
\newif\ifstillc@nverging
\def \Mess@ge #1{\ifdebug \then \message {#1} \fi}

{ %%% Things that need abnormal catcodes %%%
	\catcode `\@ = \letter
	\gdef \nodimen {\expandafter \n@dimen \the \dimen}
	\gdef \term #1 #2 #3%
	       {\edef \t@ {\the #1}%%% freeze parameter 1 (count, by value)
		\edef \t@@ {\expandafter \n@dimen \the #2\r@dian}%
				   %%% freeze parameter 2 (dimen, by value)
		\t@rm {\t@} {\t@@} {#3}%
	       }
	\gdef \t@rm #1 #2 #3%
	       {{%
		\count 0 = 0
		\dimen 0 = 1 \dimensionless@nit
		\dimen 2 = #2\relax
		\Mess@ge {Calculating term #1 of \nodimen 2}%
		\loop
		\ifnum	\count 0 < #1
		\then	\advance \count 0 by 1
			\Mess@ge {Iteration \the \count 0 \space}%
			\Multiply \dimen 0 by {\dimen 2}%
			\Mess@ge {After multiplication, term = \nodimen 0}%
			\Divide \dimen 0 by {\count 0}%
			\Mess@ge {After division, term = \nodimen 0}%
		\repeat
		\Mess@ge {Final value for term #1 of
				\nodimen 2 \space is \nodimen 0}%
		\xdef \Term {#3 = \nodimen 0 \r@dians}%
		\aftergroup \Term
	       }}
	\catcode `\p = \other
	\catcode `\t = \other
	\gdef \n@dimen #1pt{#1} %%% throw away the ``pt''
}

\def \Divide #1by #2{\divide #1 by #2} %%% just a synonym

\def \Multiply #1by #2%%% allows division of a dimen by a dimen
       {{%%% should really freeze parameter 2 (dimen, passed by value)
	\count 0 = #1\relax
	\count 2 = #2\relax
	\count 4 = 65536
	\Mess@ge {Before scaling, count 0 = \the \count 0 \space and
			count 2 = \the \count 2}%
	\ifnum	\count 0 > 32767 %%% do our best to avoid overflow
	\then	\divide \count 0 by 4
		\divide \count 4 by 4
	\else	\ifnum	\count 0 < -32767
		\then	\divide \count 0 by 4
			\divide \count 4 by 4
		\else
		\fi
	\fi
	\ifnum	\count 2 > 32767 %%% while retaining reasonable accuracy
	\then	\divide \count 2 by 4
		\divide \count 4 by 4
	\else	\ifnum	\count 2 < -32767
		\then	\divide \count 2 by 4
			\divide \count 4 by 4
		\else
		\fi
	\fi
	\multiply \count 0 by \count 2
	\divide \count 0 by \count 4
	\xdef \product {#1 = \the \count 0 \internal@nits}%
	\aftergroup \product
       }}

\def\r@duce{\ifdim\dimen0 > 90\r@dian \then   % sin(x+90) = sin(180-x)
		\multiply\dimen0 by -1
		\advance\dimen0 by 180\r@dian
		\r@duce
	    \else \ifdim\dimen0 < -90\r@dian \then  % sin(-x) = sin(360+x)
		\advance\dimen0 by 360\r@dian
		\r@duce
		\fi
	    \fi}

\def\Sine#1%
       {{%
	\dimen 0 = #1 \r@dian
	\r@duce
	\ifdim\dimen0 = -90\r@dian \then
	   \dimen4 = -1\r@dian
	   \c@mputefalse
	\fi
	\ifdim\dimen0 = 90\r@dian \then
	   \dimen4 = 1\r@dian
	   \c@mputefalse
	\fi
	\ifdim\dimen0 = 0\r@dian \then
	   \dimen4 = 0\r@dian
	   \c@mputefalse
	\fi
	\ifc@mpute \then
        	% convert degrees to radians
		\divide\dimen0 by 180
		\dimen0=3.141592654\dimen0
		\dimen 2 = 3.1415926535897963\r@dian %%% a well-known constant
		\divide\dimen 2 by 2 %%% we only deal with -pi/2 : pi/2
		\Mess@ge {Sin: calculating Sin of \nodimen 0}%
		\count 0 = 1 %%% see power-series expansion for sine
		\dimen 2 = 1 \r@dian %%% ditto
		\dimen 4 = 0 \r@dian %%% ditto
		\loop
			\ifnum	\dimen 2 = 0 %%% then we've done
			\then	\stillc@nvergingfalse
			\else	\stillc@nvergingtrue
			\fi
			\ifstillc@nverging %%% then calculate next term
			\then	\term {\count 0} {\dimen 0} {\dimen 2}%
				\advance \count 0 by 2
				\count 2 = \count 0
				\divide \count 2 by 2
				\ifodd	\count 2 %%% signs alternate
				\then	\advance \dimen 4 by \dimen 2
				\else	\advance \dimen 4 by -\dimen 2
				\fi
		\repeat
	\fi
			\xdef \sine {\nodimen 4}%
       }}

% Now the Cosine can be calculated easily by calling \Sine
\def\Cosine#1{\ifx\sine\UnDefined\edef\Savesine{\relax}\else
		             \edef\Savesine{\sine}\fi
	{\dimen0=#1\r@dian\advance\dimen0 by 90\r@dian
	 \Sine{\nodimen 0}
	 \xdef\cosine{\sine}
	 \xdef\sine{\Savesine}}}
% end of trig stuff
%%%%%%%%%%%%%%%%%%%%%%%%%%%%%%%%%%%%%%%%%%%%%%%%%%%%%%%%%%%%%%%%%%%%

\def\psdraft{
	\def\@psdraft{0}
	%\ps@typeout{draft level now is \@psdraft \space . }
}
\def\psfull{
	\def\@psdraft{100}
	%\ps@typeout{draft level now is \@psdraft \space . }
}

\psfull

\newif\if@scalefirst
\def\psscalefirst{\@scalefirsttrue}
\def\psrotatefirst{\@scalefirstfalse}
\psrotatefirst

\newif\if@draftbox
\def\psnodraftbox{
	\@draftboxfalse
}
\def\psdraftbox{
	\@draftboxtrue
}
\@draftboxtrue

\newif\if@prologfile
\newif\if@postlogfile
\def\pssilent{
	\@noisyfalse
}
\def\psnoisy{
	\@noisytrue
}
\psnoisy
%%% These are for the option list.
%%% A specification of the form a = b maps to calling \@p@@sa{b}
\newif\if@bbllx
\newif\if@bblly
\newif\if@bburx
\newif\if@bbury
\newif\if@height
\newif\if@width
\newif\if@rheight
\newif\if@rwidth
\newif\if@angle
\newif\if@clip
\newif\if@verbose
\def\@p@@sclip#1{\@cliptrue}

\newif\if@decmpr

%%% GDH 7/26/87 -- changed so that it first looks in the local directory,
%%% then in a specified global directory for the ps file.
%%% RPR 6/25/91 -- changed so that it defaults to user-supplied name if
%%% boundingbox info is specified, assuming graphic will be created by
%%% print time.
%%% TJD 10/19/91 -- added bbfile vs. file distinction, and @decmpr flag

\def\@p@@sfigure#1{\def\@p@sfile{null}\def\@p@sbbfile{null}
	        \openin1=#1.bb
		\ifeof1\closein1
	        	\openin1=\figurepath#1.bb
			\ifeof1\closein1
			        \openin1=#1
				\ifeof1\closein1%
				       \openin1=\figurepath#1
					\ifeof1
					   \ps@typeout{Error, File #1 not found}
						\if@bbllx\if@bblly
				   		\if@bburx\if@bbury
			      				\def\@p@sfile{#1}%
			      				\def\@p@sbbfile{#1}%
							\@decmprfalse
				  	   	\fi\fi\fi\fi
					\else\closein1
				    		\def\@p@sfile{\figurepath#1}%
				    		\def\@p@sbbfile{\figurepath#1}%
						\@decmprfalse
	                       		\fi%
			 	\else\closein1%
					\def\@p@sfile{#1}
					\def\@p@sbbfile{#1}
					\@decmprfalse
			 	\fi
			\else
				\def\@p@sfile{\figurepath#1}
				\def\@p@sbbfile{\figurepath#1.bb}
				\@decmprtrue
			\fi
		\else
			\def\@p@sfile{#1}
			\def\@p@sbbfile{#1.bb}
			\@decmprtrue
		\fi}

\def\@p@@sfile#1{\@p@@sfigure{#1}}

\def\@p@@sbbllx#1{
		%\ps@typeout{bbllx is #1}
		\@bbllxtrue
		\dimen100=#1
		\edef\@p@sbbllx{\number\dimen100}
}
\def\@p@@sbblly#1{
		%\ps@typeout{bblly is #1}
		\@bbllytrue
		\dimen100=#1
		\edef\@p@sbblly{\number\dimen100}
}
\def\@p@@sbburx#1{
		%\ps@typeout{bburx is #1}
		\@bburxtrue
		\dimen100=#1
		\edef\@p@sbburx{\number\dimen100}
}
\def\@p@@sbbury#1{
		%\ps@typeout{bbury is #1}
		\@bburytrue
		\dimen100=#1
		\edef\@p@sbbury{\number\dimen100}
}
\def\@p@@sheight#1{
		\@heighttrue
		\dimen100=#1
   		\edef\@p@sheight{\number\dimen100}
		%\ps@typeout{Height is \@p@sheight}
}
\def\@p@@swidth#1{
		%\ps@typeout{Width is #1}
		\@widthtrue
		\dimen100=#1
		\edef\@p@swidth{\number\dimen100}
}
\def\@p@@srheight#1{
		%\ps@typeout{Reserved height is #1}
		\@rheighttrue
		\dimen100=#1
		\edef\@p@srheight{\number\dimen100}
}
\def\@p@@srwidth#1{
		%\ps@typeout{Reserved width is #1}
		\@rwidthtrue
		\dimen100=#1
		\edef\@p@srwidth{\number\dimen100}
}
\def\@p@@sangle#1{
		%\ps@typeout{Rotation is #1}
		\@angletrue
%		\dimen100=#1
		\edef\@p@sangle{#1} %\number\dimen100}
}
\def\@p@@ssilent#1{
		\@verbosefalse
}
\def\@p@@sprolog#1{\@prologfiletrue\def\@prologfileval{#1}}
\def\@p@@spostlog#1{\@postlogfiletrue\def\@postlogfileval{#1}}
\def\@cs@name#1{\csname #1\endcsname}
\def\@setparms#1=#2,{\@cs@name{@p@@s#1}{#2}}
%
% initialize the defaults (size the size of the figure)
%
\def\ps@init@parms{
		\@bbllxfalse \@bbllyfalse
		\@bburxfalse \@bburyfalse
		\@heightfalse \@widthfalse
		\@rheightfalse \@rwidthfalse
		\def\@p@sbbllx{}\def\@p@sbblly{}
		\def\@p@sbburx{}\def\@p@sbbury{}
		\def\@p@sheight{}\def\@p@swidth{}
		\def\@p@srheight{}\def\@p@srwidth{}
		\def\@p@sangle{0}
		\def\@p@sfile{} \def\@p@sbbfile{}
		\def\@p@scost{10}
		\def\@sc{}
		\@prologfilefalse
		\@postlogfilefalse
		\@clipfalse
		\if@noisy
			\@verbosetrue
		\else
			\@verbosefalse
		\fi
}
%
% Go through the options setting things up.
%
\def\parse@ps@parms#1{
	 	\@psdo\@psfiga:=#1\do
		   {\expandafter\@setparms\@psfiga,}}
%
% Compute bb height and width
%
\newif\ifno@bb
\def\bb@missing{
	\if@verbose{
		\ps@typeout{psfig: searching \@p@sbbfile \space  for bounding box}
	}\fi
	\no@bbtrue
	\epsf@getbb{\@p@sbbfile}
        \ifno@bb \else \bb@cull\epsf@llx\epsf@lly\epsf@urx\epsf@ury\fi
}
\def\bb@cull#1#2#3#4{
	\dimen100=#1 bp\edef\@p@sbbllx{\number\dimen100}
	\dimen100=#2 bp\edef\@p@sbblly{\number\dimen100}
	\dimen100=#3 bp\edef\@p@sbburx{\number\dimen100}
	\dimen100=#4 bp\edef\@p@sbbury{\number\dimen100}
	\no@bbfalse
}
% rotate point (#1,#2) about (0,0).
% The sine and cosine of the angle are already stored in \sine and
% \cosine.  The result is placed in (\p@intvaluex, \p@intvaluey).
\newdimen\p@intvaluex
\newdimen\p@intvaluey
\def\rotate@#1#2{{\dimen0=#1 sp\dimen1=#2 sp
%            	calculate x' = x \cos\theta - y \sin\theta
		  \global\p@intvaluex=\cosine\dimen0
		  \dimen3=\sine\dimen1
		  \global\advance\p@intvaluex by -\dimen3
% 		calculate y' = x \sin\theta + y \cos\theta
		  \global\p@intvaluey=\sine\dimen0
		  \dimen3=\cosine\dimen1
		  \global\advance\p@intvaluey by \dimen3
		  }}
\def\compute@bb{
		\no@bbfalse
		\if@bbllx \else \no@bbtrue \fi
		\if@bblly \else \no@bbtrue \fi
		\if@bburx \else \no@bbtrue \fi
		\if@bbury \else \no@bbtrue \fi
		\ifno@bb \bb@missing \fi
		\ifno@bb \ps@typeout{FATAL ERROR: no bb supplied or found}
			\no-bb-error
		\fi
		%
%\ps@typeout{BB: \@p@sbbllx, \@p@sbblly, \@p@sbburx, \@p@sbbury}
%
% store height/width of original (unrotated) bounding box
		\count203=\@p@sbburx
		\count204=\@p@sbbury
		\advance\count203 by -\@p@sbbllx
		\advance\count204 by -\@p@sbblly
		\edef\ps@bbw{\number\count203}
		\edef\ps@bbh{\number\count204}
		%\ps@typeout{ psbbh = \ps@bbh, psbbw = \ps@bbw }
		\if@angle
			\Sine{\@p@sangle}\Cosine{\@p@sangle}
	        	{\dimen100=\maxdimen\xdef\r@p@sbbllx{\number\dimen100}
					    \xdef\r@p@sbblly{\number\dimen100}
			                    \xdef\r@p@sbburx{-\number\dimen100}
					    \xdef\r@p@sbbury{-\number\dimen100}}
%
% Need to rotate all four points and take the X-Y extremes of the new
% points as the new bounding box.
                        \def\minmaxtest{
			   \ifnum\number\p@intvaluex<\r@p@sbbllx
			      \xdef\r@p@sbbllx{\number\p@intvaluex}\fi
			   \ifnum\number\p@intvaluex>\r@p@sbburx
			      \xdef\r@p@sbburx{\number\p@intvaluex}\fi
			   \ifnum\number\p@intvaluey<\r@p@sbblly
			      \xdef\r@p@sbblly{\number\p@intvaluey}\fi
			   \ifnum\number\p@intvaluey>\r@p@sbbury
			      \xdef\r@p@sbbury{\number\p@intvaluey}\fi
			   }
%			lower left
			\rotate@{\@p@sbbllx}{\@p@sbblly}
			\minmaxtest
%			upper left
			\rotate@{\@p@sbbllx}{\@p@sbbury}
			\minmaxtest
%			lower right
			\rotate@{\@p@sbburx}{\@p@sbblly}
			\minmaxtest
%			upper right
			\rotate@{\@p@sbburx}{\@p@sbbury}
			\minmaxtest
			\edef\@p@sbbllx{\r@p@sbbllx}\edef\@p@sbblly{\r@p@sbblly}
			\edef\@p@sbburx{\r@p@sbburx}\edef\@p@sbbury{\r@p@sbbury}
%\ps@typeout{rotated BB: \r@p@sbbllx, \r@p@sbblly, \r@p@sbburx, \r@p@sbbury}
		\fi
		\count203=\@p@sbburx
		\count204=\@p@sbbury
		\advance\count203 by -\@p@sbbllx
		\advance\count204 by -\@p@sbblly
		\edef\@bbw{\number\count203}
		\edef\@bbh{\number\count204}
		%\ps@typeout{ bbh = \@bbh, bbw = \@bbw }
}
%
% \in@hundreds performs #1 * (#2 / #3) correct to the hundreds,
%	then leaves the result in @result
%
\def\in@hundreds#1#2#3{\count240=#2 \count241=#3
		     \count100=\count240	% 100 is first digit #2/#3
		     \divide\count100 by \count241
		     \count101=\count100
		     \multiply\count101 by \count241
		     \advance\count240 by -\count101
		     \multiply\count240 by 10
		     \count101=\count240	%101 is second digit of #2/#3
		     \divide\count101 by \count241
		     \count102=\count101
		     \multiply\count102 by \count241
		     \advance\count240 by -\count102
		     \multiply\count240 by 10
		     \count102=\count240	% 102 is the third digit
		     \divide\count102 by \count241
		     \count200=#1\count205=0
		     \count201=\count200
			\multiply\count201 by \count100
		 	\advance\count205 by \count201
		     \count201=\count200
			\divide\count201 by 10
			\multiply\count201 by \count101
			\advance\count205 by \count201
		     \count201=\count200
			\divide\count201 by 100
			\multiply\count201 by \count102
			\advance\count205 by \count201
		     \edef\@result{\number\count205}
}
\def\compute@wfromh{
		% computing : width = height * (bbw / bbh)
		\in@hundreds{\@p@sheight}{\@bbw}{\@bbh}
		%\ps@typeout{ \@p@sheight * \@bbw / \@bbh, = \@result }
		\edef\@p@swidth{\@result}
		%\ps@typeout{w from h: width is \@p@swidth}
}
\def\compute@hfromw{
		% computing : height = width * (bbh / bbw)
	        \in@hundreds{\@p@swidth}{\@bbh}{\@bbw}
		%\ps@typeout{ \@p@swidth * \@bbh / \@bbw = \@result }
		\edef\@p@sheight{\@result}
		%\ps@typeout{h from w : height is \@p@sheight}
}
\def\compute@handw{
		\if@height
			\if@width
			\else
				\compute@wfromh
			\fi
		\else
			\if@width
				\compute@hfromw
			\else
				\edef\@p@sheight{\@bbh}
				\edef\@p@swidth{\@bbw}
			\fi
		\fi
}
\def\compute@resv{
		\if@rheight \else \edef\@p@srheight{\@p@sheight} \fi
		\if@rwidth \else \edef\@p@srwidth{\@p@swidth} \fi
		%\ps@typeout{rheight = \@p@srheight, rwidth = \@p@srwidth}
}
%
% Compute any missing values
\def\compute@sizes{
	\compute@bb
	\if@scalefirst\if@angle
% at this point the bounding box has been adjsuted correctly for
% rotation.  PSFIG does all of its scaling using \@bbh and \@bbw.  If
% a width= or height= was specified along with \psscalefirst, then the
% width=/height= value needs to be adjusted to match the new (rotated)
% bounding box size (specifed in \@bbw and \@bbh).
%    \ps@bbw       width=
%    -------  =  ----------
%    \@bbw       new width=
% so `new width=' = (width= * \@bbw) / \ps@bbw; where \ps@bbw is the
% width of the original (unrotated) bounding box.
	\if@width
	   \in@hundreds{\@p@swidth}{\@bbw}{\ps@bbw}
	   \edef\@p@swidth{\@result}
	\fi
	\if@height
	   \in@hundreds{\@p@sheight}{\@bbh}{\ps@bbh}
	   \edef\@p@sheight{\@result}
	\fi
	\fi\fi
	\compute@handw
	\compute@resv}

%
% \psfig
% usage : \psfig{file=, height=, width=, bbllx=, bblly=, bburx=, bbury=,
%			rheight=, rwidth=, clip=}
%
% "clip=" is a switch and takes no value, but the `=' must be present.
\def\psfig#1{\vbox {
	% do a zero width hard space so that a single
	% \psfig in a centering enviornment will behave nicely
	%{\setbox0=\hbox{\ }\ \hskip-\wd0}
	%
	\ps@init@parms
	\parse@ps@parms{#1}
	\compute@sizes
	\ifnum\@p@scost<\@psdraft{
		\special{ps::[begin] 	\@p@swidth \space \@p@sheight \space
				\@p@sbbllx \space \@p@sbblly \space
				\@p@sbburx \space \@p@sbbury \space
				startTexFig \space }
		\if@angle
			\special {ps:: \@p@sangle \space rotate \space}
		\fi
		\if@clip{
			\if@verbose{
				\ps@typeout{(clip)}
			}\fi
			\special{ps:: doclip \space }
		}\fi
		\if@prologfile
		    \special{ps: plotfile \@prologfileval \space } \fi
		\if@decmpr{
			\if@verbose{
				\ps@typeout{psfig: including \@p@sfile.Z \space }
			}\fi
			\special{ps: plotfile "`zcat \@p@sfile.Z" \space }
		}\else{
			\if@verbose{
				\ps@typeout{psfig: including \@p@sfile \space }
			}\fi
			\special{ps: plotfile \@p@sfile \space }
		}\fi
		\if@postlogfile
		    \special{ps: plotfile \@postlogfileval \space } \fi
		\special{ps::[end] endTexFig \space }
		% Create the vbox to reserve the space for the figure
		\vbox to \@p@srheight true sp{
			\hbox to \@p@srwidth true sp{
				\hss
			}
		\vss
		}
	}\else{
		% draft figure, just reserve the space and print the
		% path name.
		\if@draftbox{
			% Verbose draft: print file name in box
			\hbox{\frame{\vbox to \@p@srheight true sp{
			\vss
			\hbox to \@p@srwidth true sp{ \hss \@p@sfile \hss }
			\vss
			}}}
		}\else{
			% Non-verbose draft
			\vbox to \@p@srheight true sp{
			\vss
			\hbox to \@p@srwidth true sp{\hss}
			\vss
			}
		}\fi

	}\fi
}}
\psfigRestoreAt

\def\caphead#1{\noindent\parshape 1 2cm 14cm{\it\bf#1}}
\def\caption#1{\noindent\parshape 1 2cm 14cm{\it#1}}
\def\fheight{12cm}
\def\fwidth{17cm}
\def\K{{\rm K}}

\vfill

\goodbreak
\psfig{figure=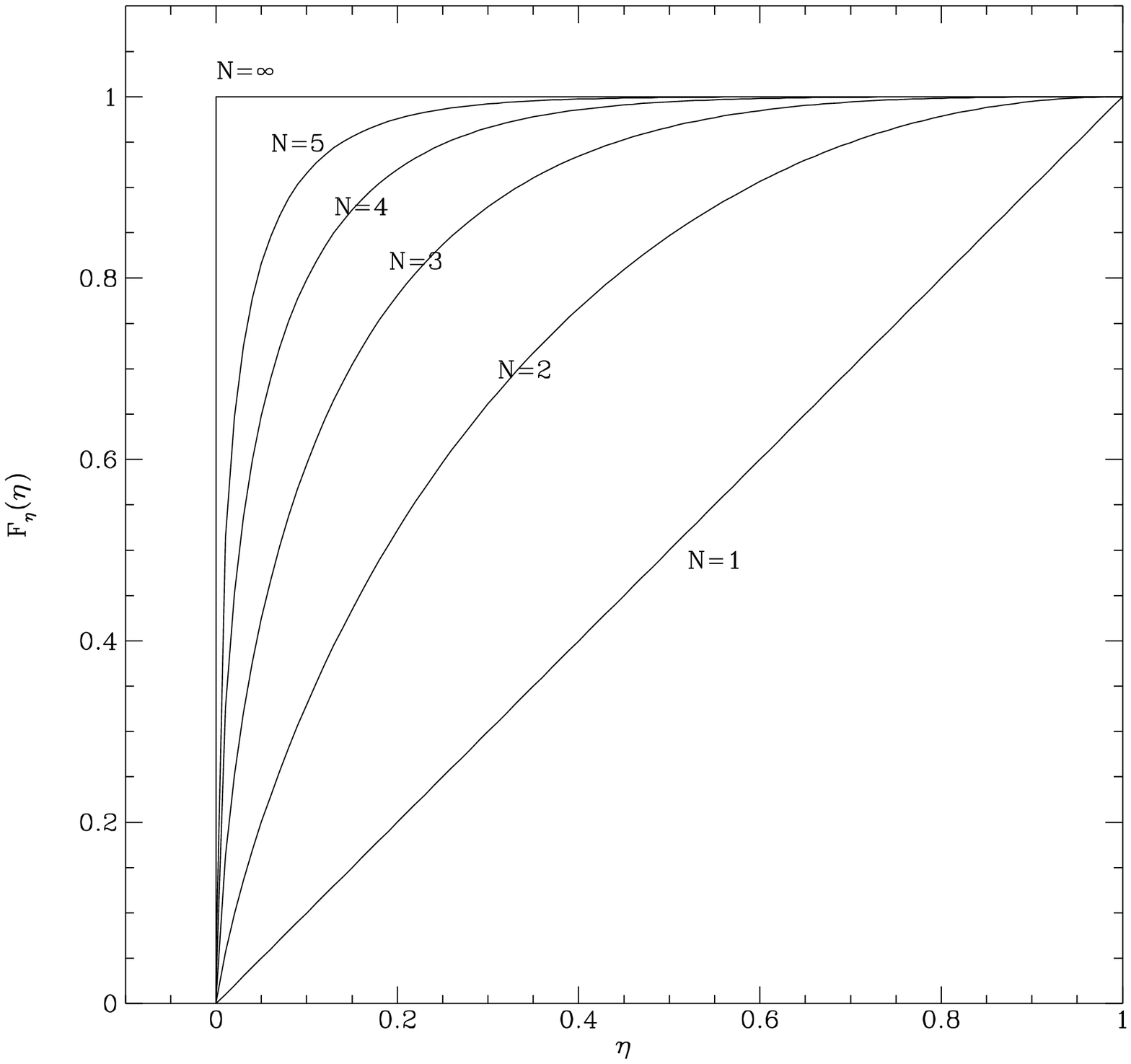,width=\fwidth,height=\fheight,prolog=/apps/tex82/common/macros/mac.pro}
\nobreak
\caphead{Figure 1. The function $F_{\eta}$.}

\caption{
The cumulative probability distribution for the goodness-of-fit parameter
$\eta$ is plotted for a few different $n$-values.
}
\vfill

\def\fheight{12cm}
\def\fwidth{17cm}

\goodbreak
\psfig{figure=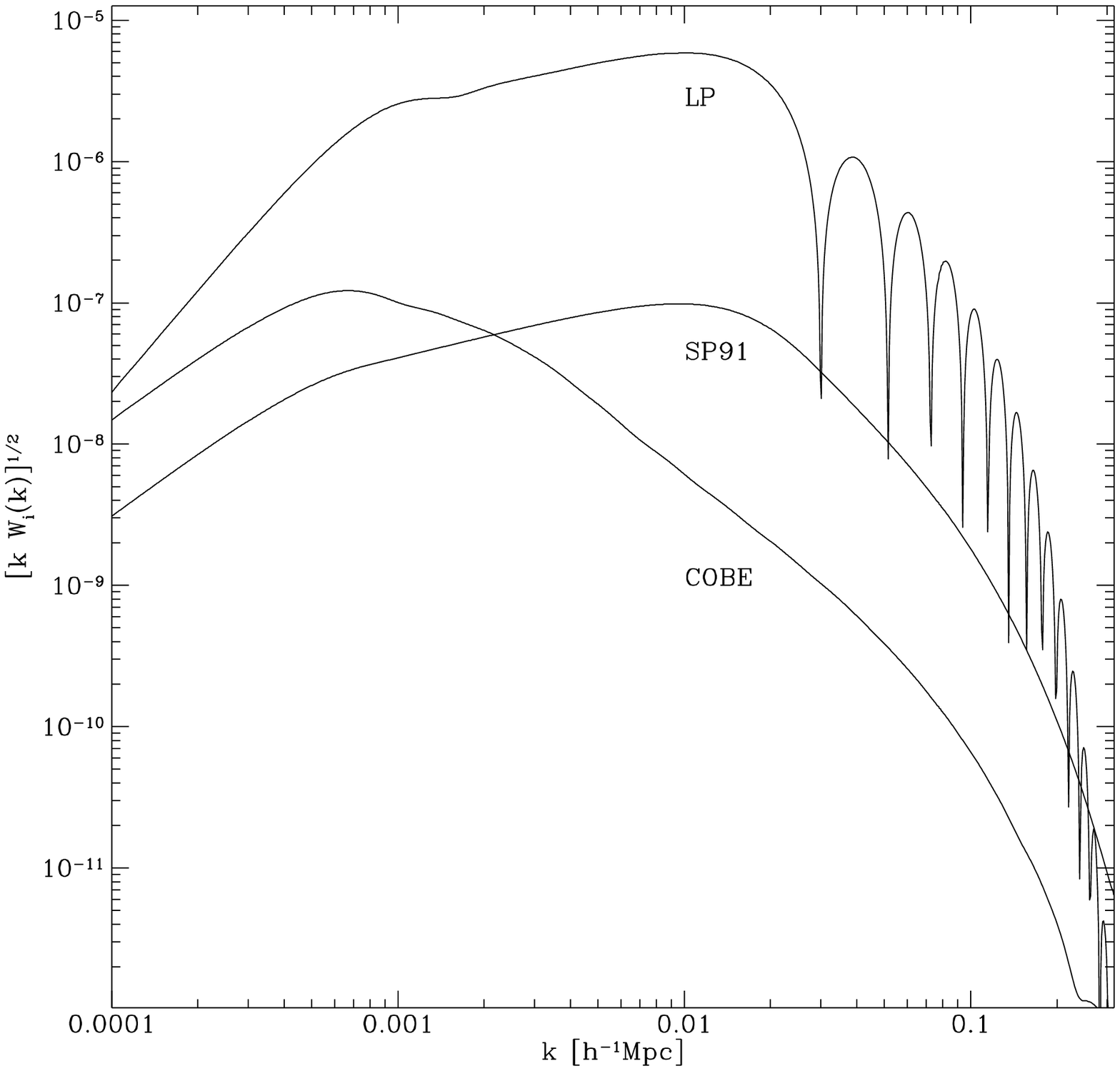,width=\fwidth,height=\fheight,prolog=/apps/tex82/common/macros/mac.pro}
\nobreak
\caphead{Figure 2. Window Functions.}

\caption{
The window functions of the Lauer-Postman
bulk flow measurement (LP), the
South Pole 1991 nine point scan (SP91),
and the COBE DMR $10^{\circ}$ pixel {r.m.s.}
measurement (COBE) are plotted as a function of
comoving wavenumber $k$.
}
\vfill

\goodbreak
\psfig{figure=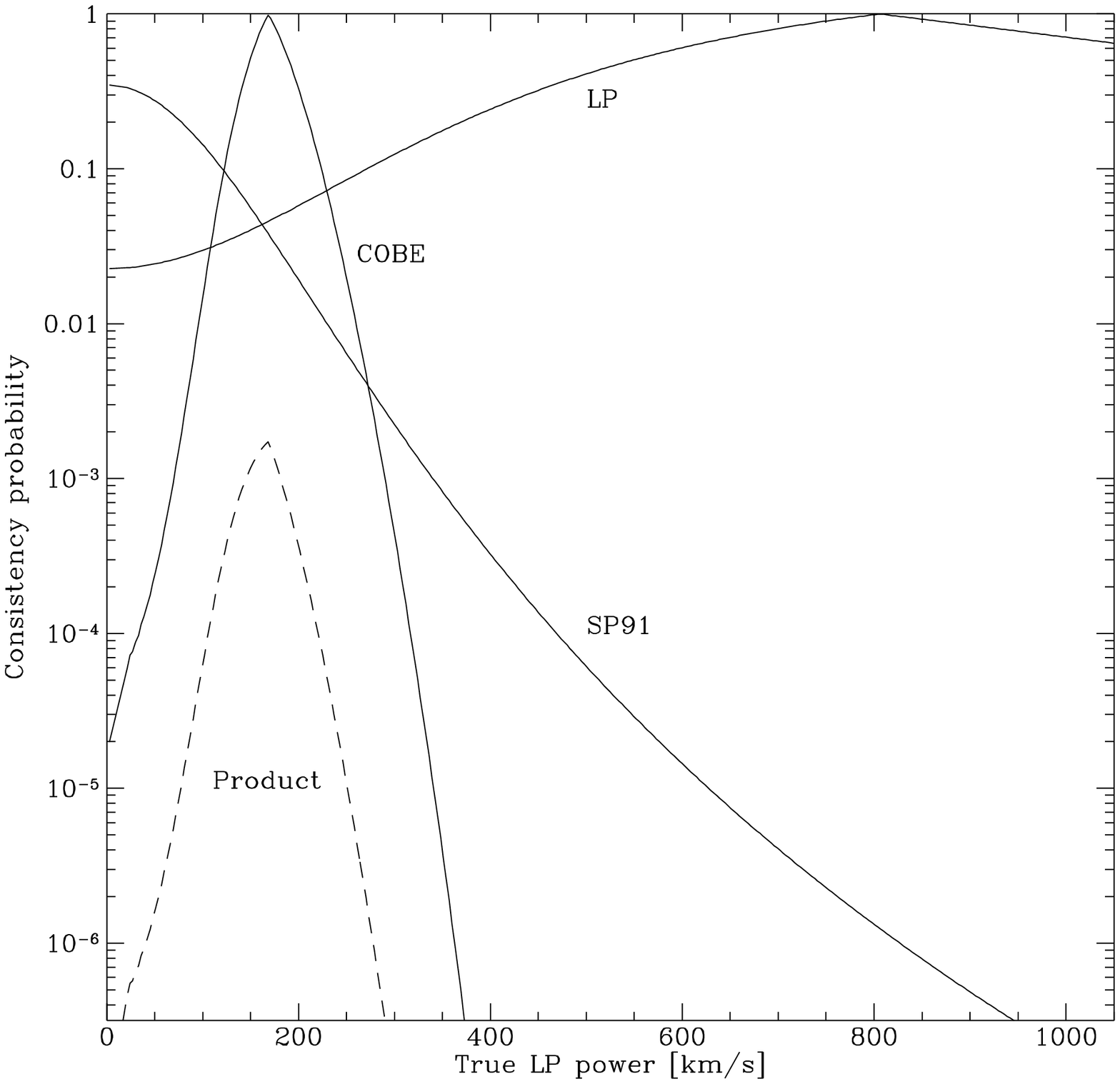,width=\fwidth,height=\fheight,prolog=/apps/tex82/common/macros/mac.pro}
\nobreak
\caphead{Figure 3. Consistency Probabilities}

\caption{
The probability that the the Lauer-Postman
bulk flow measurement (LP), the COBE DMR experiment and the
South Pole 1991 nine point scan (SP91)
are consistent with CDM is plotted as a function of the normalization
of the power spectrum. The normalization is expressed in terms of
the expected bulk flow in a LP measurement.
The dashed line is the product of these three probabilities, and takes a
maximum for a normalization corresponding to 168 km/s.
}
\vfill

\goodbreak
\psfig{figure=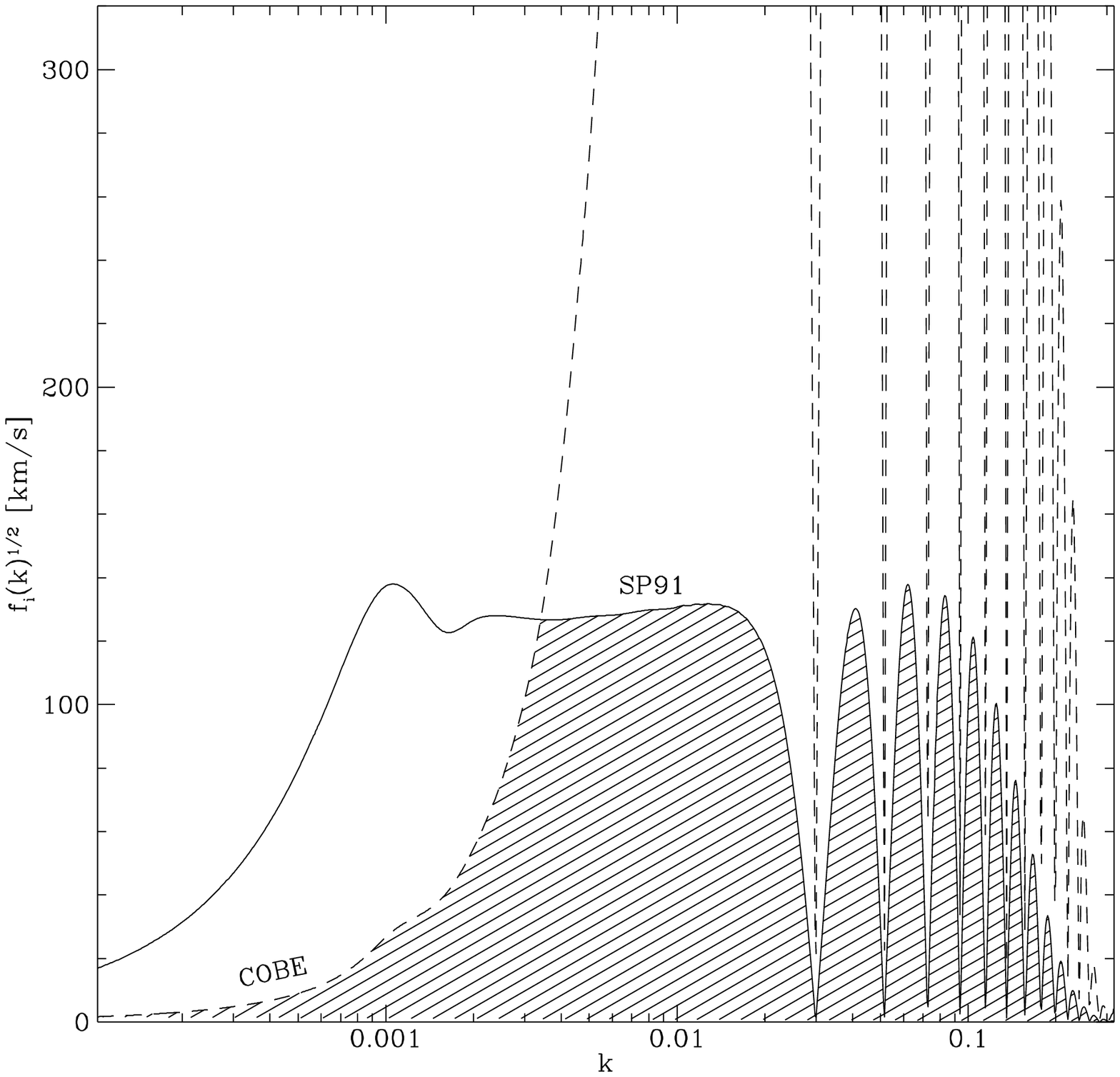,width=\fwidth,height=\fheight,prolog=/apps/tex82/common/macros/mac.pro}
\caphead{Figure 4. The best places to hide power}

\caption{
The functions $f_{sp}$ (solid line) and $f{cobe}$ (dashed line)
are plotted as a
function of wavenumber $k$.
The shaded region, {\it i.e.} the area lying beneath both curves,
constitutes the LP bulk flows that would be consistent with
the SP91 and COBE experiments using the $N=1$ constraints only, when
the power spectrum is a single delta function located at $k$.
}
\vfill

\end